\documentclass[lettersize,journal]{IEEEtran}
\usepackage{cite}
\pdfoutput=1
\usepackage{url}
\usepackage{graphicx}
\usepackage{caption}
\usepackage{subcaption}
\usepackage{amsmath,amsfonts,bm}
\usepackage{algorithm}
\usepackage{algorithmic}
\usepackage{array}
\usepackage{amssymb}
\usepackage{multirow}
\usepackage{multicol}

\floatname{algorithm}{Algorithm}
\usepackage{hyperref}
\begin{document}
\title{Raptor Encoding for Low-Latency Concurrent Multi-PDU Session Transmission with Security Consideration in B5G Edge Network}
\author{Zhongfu Guo, Xinsheng Ji, Wei You, Mingyan Xu, Yu Zhao, Zhimo Cheng, Deqiang Zhou
\thanks{Manuscript created October, 2022; Project supported by the National Key Research and Development Program of China (Nos. 2022YFB2902204 and 2020YFB1806607)and the National Natural Science Foundation of China (No. 61941114).
		
Zhongfu Guo, Wei You, Mingyan Xu, Yu Zhao, Zhimo Cheng, Deqiang Zhou are with the Department of next-generation mobile communication and cyber space security, Information Engineering University, Zhengzhou 450001, China (e-mail: ndscgzf@163.com).
		
Xinsheng Ji is with the National Digital Switching System Engineering and Technological Research and Development Center, Zhengzhou 450000,China and also with the Purple Mountain Laboratories: Networking, Communications and security, Nanjing 211111, China.}}
\markboth{Preprint submitted to IET}%
%\markboth{IET communications , ~Vol.~ , No.~ ,  ~  }%
{}

\maketitle

\begin{abstract}
	In B5G edge networks, end-to-end low-latency and high-reliability transmissions between edge computing nodes and terminal devices are essential. This paper investigates the queue-aware coding scheduling transmission of randomly arriving data packets, taking into account potential eavesdroppers in edge networks. To address these concerns, we introduce SCLER, a Protocol Data Units (PDU) Raptor-encoded multi-path transmission method that overcomes the challenges of a larger attack surface in Concurrent Multipath Transfer (CMT), excessive delay due to asymmetric delay\&bandwidth, and lack of interaction among PDU session bearers. We propose a secure and reliable transmission scheme based on Raptor encoding and distribution that incorporates a queue length-aware encoding strategy. This strategy is modeled using Constrained Markov Decision Process (CMDP), and we solve the constraint optimization problem of optimal decision-making based on a threshold strategy. Numerical results indicate that SCLER effectively reduces data leakage risks while achieving the optimal balance between delay and reliability, thereby ensuring data security. Importantly, the proposed system is compatible with current mobile networks and demonstrates practical applicability.
\end{abstract}	

\begin{IEEEkeywords}
	B5G Core Network; Reliable Communication; Concurrent Multipath Transfer; Raptor Codes; Security.
\end{IEEEkeywords}

%%%%%%%%%%%%%%%%%%%%%%%%%%%%%%%%%%%%%%%%%%
\section{Introduction}
In the B5G era\cite{3gpp.21.917}, mobile communication services such as smart cities\cite{javed2022future} and telemedicine  have increasing demands in terms of low-latency and high-reliability transmission\cite{navarro2020survey, li20185g}. B5G edge networks\cite{adhikari20226g}, powered by B5G mobile networks\cite{zhang2022mobile} and edge computing\cite{hassan2019edge}, have been proposed as a promising solution. Edge computing nodes (DN) provide customized services for users\cite{hui2021secure}, while the B5G mobile networks offer reliable transmission. Due to the highly customized characteristics of edge service flows, such as medical, smart home, and unmanned factories, data leakage can have serious consequences\cite{tedeschi2019edge}. Therefore, secure data transmission is as crucial as low-latency and high-reliability for edge networks\cite{yoshizawa2019overview}.

To address these pressing demand\cite{chen2019physical,hamamreh2017ofdm}, researchers have conducted extensive studies on the physical layer\cite{lien20175g} and MAC-layer\cite{shrivastava20225g}, with a focus on point-to-point security through methods such as pilot-assisted techniques, physical layer encoding, and interface diversity\cite{xie2022optimizing,xu2021quantum,farhat2021secure}. These efforts aim to enhance the transmission performance from UE to gNB and require infrastructure renovation accordingly.

\begin{figure}[!t]
	\centering
	\includegraphics[width=3.5in]{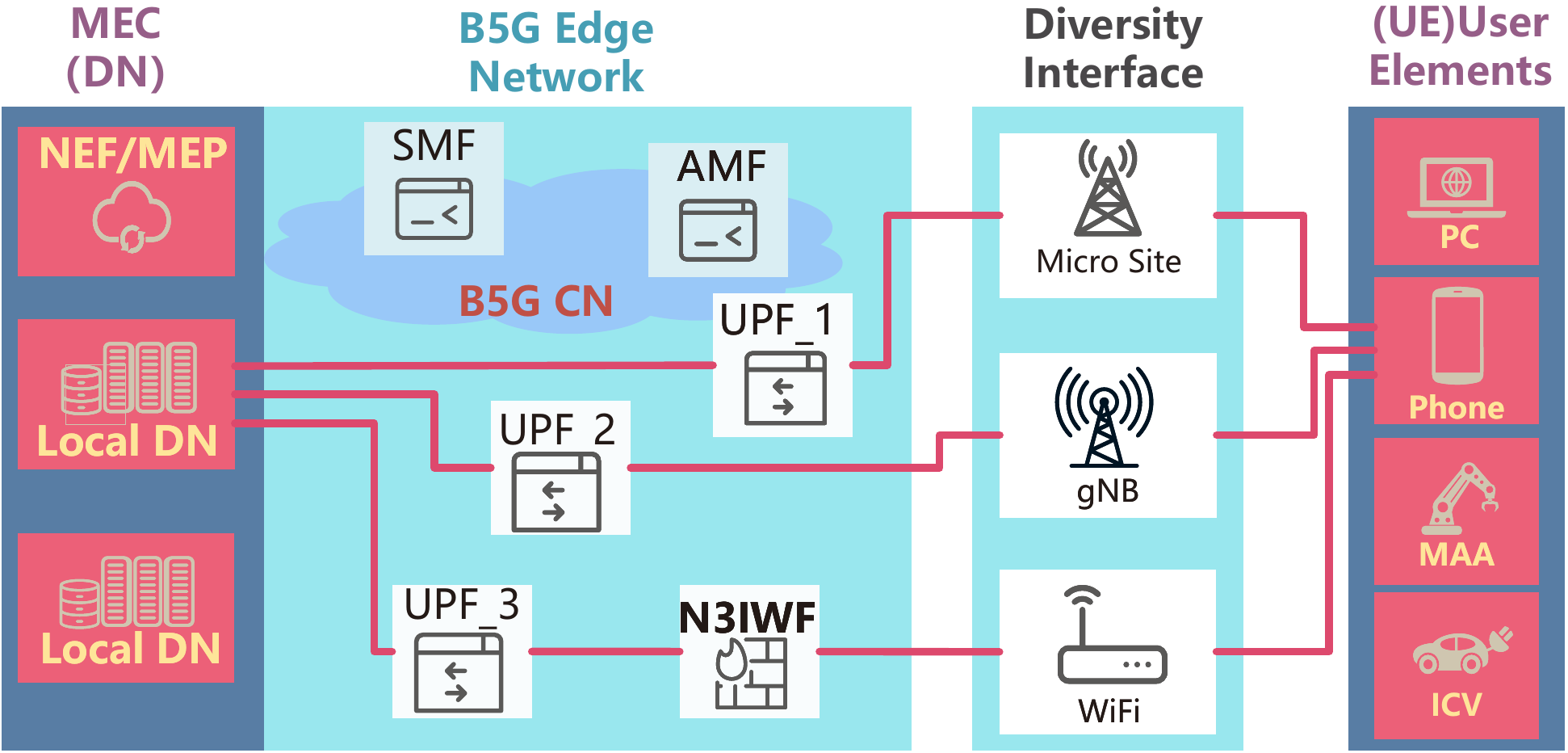}
	\caption{Enhanced Core Network for end-to-end redundant user plane paths using Diversity Interface PDU sessions.}
	\label{SCLER-FIG1}
\end{figure}

The Third Generation Partnership Project (3GPP) has proposed the Enhanced Core Network (ECN) in \cite[clause 5.33]{3gpp.23.501}. As illustrated in Fig.\ref{SCLER-FIG1}, ECN carries the end-to-end service flow between UE and DN with multiple PDU sessions, which improves reliability through redundant transmission. A multi-path transmission protocol stack is deployed on both the client (UE) and server (DN) sides. Additionally, the high-layer (above the IP layer \cite{3gpp.23.501}) splits and aggregates the service flows. However, the implementation of Concurrent Multipath Transfer (CMT) over erasure channels in Diversity Interface Edge Network (DIEN) faces several technical challenges. For example, CMT has a larger attack surface than a single session transmission \cite{yoshizawa2019overview}, and asymmetric latency and bandwidth can lead to excessive delay \cite{chen2018ultra, suer2019multi, li2016multipath}, moreover, it is hard for the PDU session bearer to interact during the transmission phase \cite{ha2019support}. For instance, Wang \textit{et al.} \cite{wang2022secure} have proposed a comprehensive method that considers both reliability and security. Security-based methods such as network coding \cite{cohen2021network}, Cross-locking \cite{huang2022cross}, etc. based on link interaction are not applicable. Addressing these challenges is essential for CMT in DIEN.

The Fountain code \cite{luby2002lt} is suitable for optimizing CMT transmission\cite{he2021delay}, due to its characteristic of being able to recover the entire source block with any set of Coded Packet($Pc$) with cardinality equal to or slightly more than the number of Source Packet($Ps$). Cui \textit{et al.}\cite{cui2014fmtcp} studied a Raptor-based\cite{luby2007raptor} CMT protocol, which improves scheduling flexibility and simplifies the signaling overhead of CMT management. Packet-level encoding can improve robustness; longer block lengths are typically used to ensure the reliability of the Fountain Code \cite{abbas2019performance}. 
From a security perspective, Fountain $Pc$ is a non-systematic code obtained by randomly selecting and XOR-ing $Ps$, making it challenging for eavesdroppers\footnote{Eavesdroppers generally refer to anyone who can attempt data theft at any layer, from the physical layer to the application layer.} to obtain private information from leaked $Pc$\cite{jain2020rateless, yi2014achieving}. Therefore, the number of $Pc$ transmissions on different sessions should be selected carefully to ensure both reliability and security are considered comprehensively.

Careful selection of the appropriate code length is crucial to achieving reliable and low-latency\cite{GUO2023109716}.   He \textit{et al.}\cite{he2021delay} added the RCDC sublayer of Raptor code to the MAC layer to optimize downlink transmission delay of dual base stations.  Kwon \textit{et al.}\cite{kwon2014mpmtp} added the MPMTP sublayer to the TCP layer to adjust Raptor encoding parameters to alleviate queue head congestion and transmission quality degradation.  Nielsen \textit{et al.}\cite{nielsen2017ultra} formulated an optimization problem to find payload allocation weights that maximize reliability at specific target latency values. These studies focus on the air interface of the mobile network, which requires upgrading the network infrastructure. Our research goal is to add security considerations and provide end-to-end low-latency reliable transmission for DIEN at the PDU session layer while considering the queuing delay caused by the random arrival of data packets, which is closer to real-world application scenarios.

This paper proposes SCLER\footnote{SCLER is short for Secure Low-Latency Concurrent transmission with Raptor encoding in Edge network, pronounced as "skler".}, a transmission method for Raptor coded packets $Pc$ at the PDU session layer. The goal is to provide reliable and low-latency transmission while taking into account data security. To address queuing delays that may result from the accumulation of randomly arriving data packets, SCLER incorporates a queue-aware variable-length encoder. Considering longer block lengths of $Pc$ can provide reliability and encoding efficiency, they also result in increased transmission and queuing delays. Therefore, SCLER trades off reliability and delay through variable block-length encoding.
To ensure data security, this paper considers possible randomly distributed non-cooperative passive attackers.
SCLER distributes $Pc$ to multiple PDU sessions based on their weight, which does not exceed the safety threshold,
the number of $Pc$ sent is determined by a measuring function to ensure reliable data transmission while minimizing information leakage. 
We study the strategy of Raptor CMT for DIEN based on Constrained Markov Decision Process($CMDP$) and obtain a security, delay-optimized, reliable transmission by ${\emph{Linear\  Programming}}$. The contributions of this paper are:

\begin{figure*}[!t]
	\centering
	\includegraphics[width=7in]{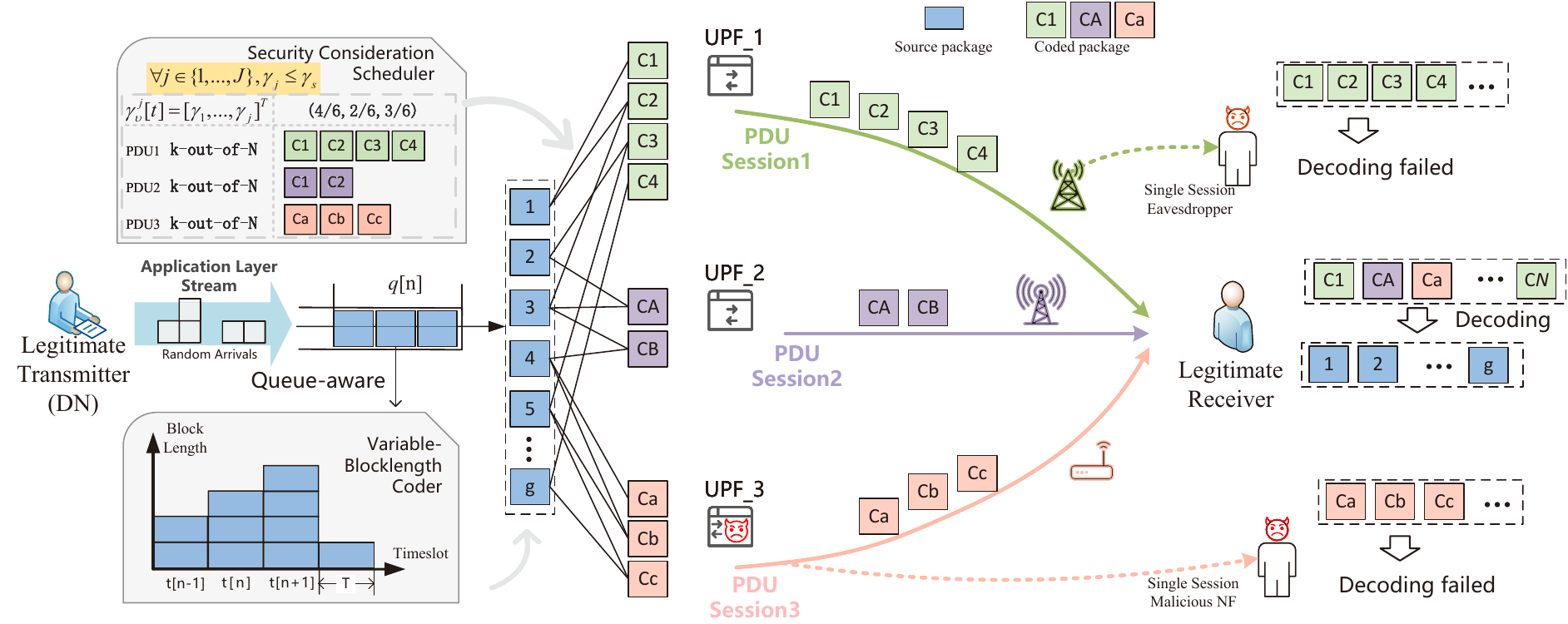}
	\caption{System Diagram.}
	\label{SCLER-FIG2}
\end{figure*} 

\begin{itemize}
	\item{Proposing SCLER, the first raptor-based security transmission scheme based on ENC implementation. SCLER considers the random distribution of any number of non-cooperative passive attackers, and makes it impossible for a single session to steal valid information.}
	
	\item{A queue-aware variable block length encoding scheme is designed, which models the optimal CMT strategies as CMDP. Feature analysis is used to simplify the LP problem into a threshold decision problem, allowing for the optimal delay-reliability trade-off to be obtained.}
	
	\item{The entire system is designed at the transport layer, ensuring compatibility with current mobile networks. This design improves service experience without requiring modifications to the core network infrastructure.}
\end{itemize}

The remaining sections of this paper are organized as follows. Section II provides a brief review of related research. Section III presents the design of a diversity interface PDU session establishment procedure for B5G networks, which serves as the B5G SCLER preliminary work. Section IV describes the system model. In Section V, we demonstrate the delay-reliability performance of SCLER based on the $CMDP$ and analyze the optimal delay-reliability trade-off. Section VI presents numerical results. Finally, we conclude the paper and propose future research directions in Section VII.
The important variables used throughout this paper are summarized in Table \ref{tab1}.

\begin{table}
	\begin{center}
		\caption{Basic Notations.}
		\label{tab1}
		\begin{tabular}{| l | l |}
			\hline
			\textbf{	Symbol }	& \textbf{ Definition}	\\
			\hline
			$Ps$,\;$Pc$ & the source packets,\;coded packets\\
			\hline
			${\mathbb{P}}$,\;${\mathbb{E}}$ & the probability, expectation value\\
			%		$\varepsilon$ & the overhead of encoding\\
			%		\hline
			%		$R$,\;$\gamma$ & the code-rate, inverse of the code-rate\\
			\hline
			$T$ &the time span of timeslot\\
			\hline
			$\Omega (x)$&  the degree distribution\\
			\hline
			$g$ & the block-length of codes\\
			\hline
			$\gamma_j$,$\gamma _\upsilon ^j$ & the scheduling weight for $j$th-PDU session,  \\
			&the weight vector $\gamma _\upsilon ^j = {[{\gamma _1},...,{\gamma _j}]^T}$\\
			\hline
			$g[t]$,$\gamma _\upsilon ^j[t]$& the encoding action within t-th timeslot\\
			\hline
			${d^{{\rm{th}}}}$&
			The constraints of delay.\\
			\hline
			$B$,$\Gamma _\upsilon ^j$&the upper bounds of $g[t]$,parent set of $\gamma _\upsilon ^j[t]$\\
			\hline
			$\rm{Z}$&the transmitting side queue buffer size\\
			\hline
			$\phi$& the size of each data packet \\
			\hline
			$\varepsilon _\upsilon ^j = [{\varepsilon _1},...,{\varepsilon _j}]$ & the erasure probability of $j$ sessions\\
			\hline
			$\Lambda[t]$,\ $\boldsymbol{\lambda}=$&the number $Ps$ arrivals within $t$-th timeslot  \\
			${[{\lambda _0},{\lambda _1},...,{\lambda _N}]}$, &the probability distribution of $\Lambda[t]=n$\\
			\hline
			${N_L}$ & the number of PDU sessions established in DIEN\\
			\hline		
		\end{tabular}
	\end{center}
\end{table}

\section{Related Work}
The mobile network is a widely accessible open network space that facilitates connection capabilities through multiple interface access.  It provides low-latency and high-reliability transmission for edge computing nodes and terminals, which is a fundamental service capability of the edge network.  As users subscribe to edge computing, communication security is a significant performance indicator that raises concerns.  This section presents the essential performance indicators for edge networks and reviews relevant studies before discussing our insights based on related research.

%系统指标最后说吧
To this end, SCLER considers the following critical aspects:

\textbf{Latency}($L$), which refers to the time from the arrival of the application layer $Ps$ to the completion of decoding, including the queuing delay $Dq$ of buffer queue accumulation at the sending end, the data encoding delay $Dec$, the data transmission delay $Dt$, and the data decoding delay $Ddc$. We consider $Ddc$ as the sum of transmission and decoding time\footnote{We assume that the calculation process of encoding and decoding introduces no time overhead.}. Therefore, we have $L=Dq+Ddc$.

\textbf{Reliability}($r$).  The system should be robust against PDU erasure channel, which we regard as the $j$th PDU session with a deletion probability of $\varepsilon _\upsilon ^j$. We encode $g$ $Ps$ (source data packets), generate and transmit $\bar g$ $Pc$ (encoded data packets), and regard the peer's completion of decoding without requesting to continue sending as reliable transmission. We can achieve high reliability transmission by adjusting $\bar g$.

\textbf{Security}, which measures the robustness of edge networks against passive attackers (such as eavesdroppers or network function\footnote{The data plane network function in the edge network mainly refers to the UPF(User Plane Function\cite{3gpp.29.244}), as shown in Fig.\ref{SCLER-FIG1} \& Fig.\ref{SCLER-FIG2}.} data thieves). We aim to ensure that data leaked by the DIEN in the presence of passive attackers cannot be deciphered into usable information.
Even if there are $N_A$ attackers who cooperate with each other and have eroded $L$ PDU sessions, the total amount of intercepted $Pc$ is $U = \sum\nolimits_{j=1}^L {A_j}$\footnote{The information carried by different nodes of a single PDU session is the same}. Based on $U$, attackers can decode the amount of $Ps$, which is $S$. We set the threshold $\gamma_s$ so that $S/g \leq \gamma_s$. That is, maintaining security against passive attackers of degree $L$\footnote{We are focusing on $\gamma_s$, which is directly correlated to the amount of leaked information. For the sake of simplification, we are only considering non-cooperative attackers, with L set to 1.}. A smaller $\gamma _s$ value indicates a higher security system.

%这一段是面对攻击的可能方法
There is a wealth of research on methods that address one or more of the goals outlined above. Encryption approaches like those presented in \cite{jadin2017securing} and \cite{apiecionek2015multi} necessitate the agreement of authentication and key distribution mechanisms in advance, which may not be appropriate for edge network users' dynamic mobile environment. Dynamic routing strategies, like the secure multi-path transmission algorithm based on fountain codes proposed in \cite{liu2022secure}, have been shown to improve security and reduce communication delays. However, it may result in some transmission capacity being wasted when predicting the eavesdropper's location. In terms of secure transmission methods based on encoding, Yi \textit{et al.} \cite{yi2014achieving} has conducted related research on the physical layer, Noura \textit {et al.}\cite{noura2022network} has proposed a secure transmission method based on network coding, and Huang \textit{et al.}\cite{huang2022cross} has proposed a secure transmission method based on cross-layer multi-path joint coding. The benefits of CMT in preventing data leakage have been demonstrated by \cite{yang2001improving}. Singh \textit{et al.}\cite{singh2015survey} proved that the transport layer is the most suitable location for providing an end-to-end connection of an application over disjoint paths. Therefore, we consider designing CMT at the transport layer.

%喷泉码背景，安全优势
Fountain codes \cite{luby2002lt} are commonly used in CMT to address issues such as asymmetric transmission delay and signaling overhead reduction \cite{cui2014fmtcp}. These codes have a variable rate between 0 and 1, and their coding method is characterized by randomness \cite{chen2018ultra, suer2019multi}. Fountain codes divide the data into multiple packets \{$P_s$\}, and then generate a new coded packet ($P_c$) by performing random selection and XOR operations on each packet. The generated $P_c$ packets are independent and equally significant, and the decoding rate is determined by the number of received symbols, regardless of the symbol types. Due to the randomness, non-systematic nature and lack of a fixed rate of fountain codes, each encoded packet contains a part of the original data, but it is inadequate to restore the original data. Secure coding transmission mechanisms can be developed based on fountain codes \cite{yi2014achieving}.

%问题分析
The design of CMT at the transport layer is a suitable approach for achieving the goals of the paper, but several issues need to be considered. Firstly, data split transmission can compromise security, so a comprehensive approach that considers both reliability and security is necessary\cite{yoshizawa2019overview}. Secondly, the delay caused by randomly arriving data packets and asymmetric multiplex transmission needs to be minimized\cite{chen2018ultra,suer2019multi}. Finally, the inability of PDU session bearers to interact during transmission phase limits the applicability of dynamic routing and network coding technologies\cite{kuo2008dynamic,cohen2021network}. Therefore, the proposed approach should be able to defend against randomly distributed non-cooperative passive attackers. The paper has identified these key points and will provide a detailed description of the preliminary work in the next section.

%%%%%%%%%%%%%%%%%%%%%%%%%%%%%%%%%%%%%%%%%%

\section{B5G SCLER Preliminary}
A major development trend of the beyond 5G core network is to support more complex services through cloud-network integration and distributed collaboration and provide customized services for each UE. This section will discuss in detail the intrusion probability of multiple PDUs and B5G RCLER preliminary work.

\subsection{Edge Network Intrusion Probability for Multiple PDU Sessions}
Physically isolated multi-access multi-PDU session edge computing networks can provide security benefits. In this subsection, we analyze the security of edge networks in the B5G era and explore the relationship between multiple disjoint PDU sessions and the probability of intrusion. Specifically, when there are ${N_L}$ disjoint PDU sessions, we investigate the probability that all channels are eavesdropped, i.e., the probability ${\mathbb{P}} [{N_L}]$ that all ${N_L}$ PDU sessions are attacked.
\begin{equation}\label{Intrusion-Probability}
	{\mathbb{P}} ({N_L}) = \prod\nolimits_{j = 1}^{{N_L}} {[1 - \prod\nolimits_{i = 1}^{{N_j}} {(1 - p_j^i)} ]}
\end{equation}
Where $p_j^i$ represents the probability that the $i$-th node on the $j$-th PDU session is eavesdropped, and ${{N_j}}$ represents the number of nodes contained in the j-th PDU session (including the base station(gNB), UPF, and UPF-A).
Identify all possible physically isolated and disjoint PDU sessions between the UE and the DN. Given a predetermined threshold value $\vartheta $, we require the connection intrusion probability $P({N_L})$ to be less than or equal to $\vartheta $ (denoted as $P({N_L}) \le \vartheta $). The value of ${N_L}$can be determined accordingly based on the required $\vartheta $. 
In the following subsection, we will describe the method for establishment the aforementioned PDU session, which serves as a fundamental requirement for the SCLER to operate effectively in the B5G core network.

\begin{figure}[!t]
	\centering
	\includegraphics[width=3.5in]{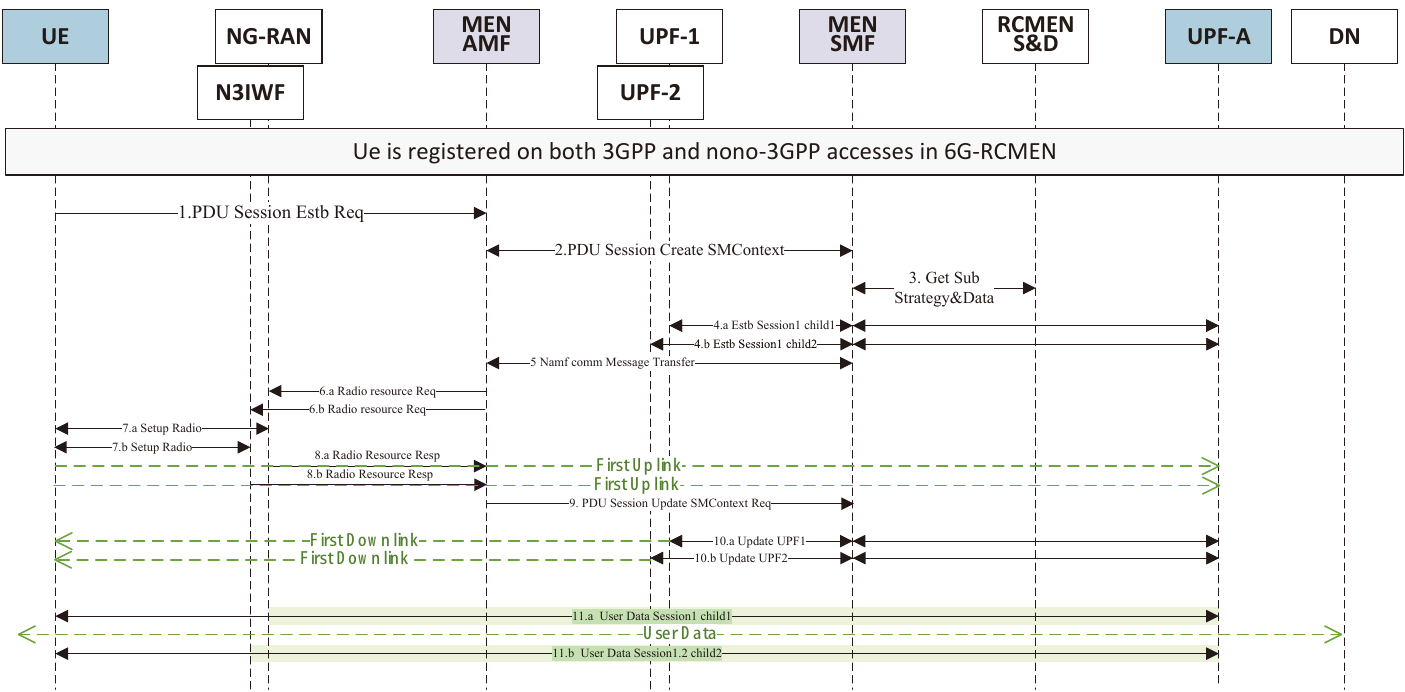}
	\caption{UE-requested physically isolated multi-access PDU session establishment procedure for B5G. (See Appendix A)}
	\label{SCLER_fig3}
\end{figure}

\subsection{Physically Isolated Multi-Access PDU Session Establish procedure}
The B5G core network we envision still follows the 5G control and bearer separation design. There are two planes, Control Plane (CP) and User Plane (UP). The core network provides PDU connectivity services between UE and DN to realize the exchange of PDUs. UP transport user PDUs. CP control manages the behavior of the entire core networks based on signaling. CP functional entities are divided into different Network Functions (NF). The User Plane Function (UPF) mainly supports the routing and forwarding of UE service data, Session Management Function (SMF) is responsible for UE session management and manages UPFs, Access and Mobility Management Function (AMF) is responsible for UE access. The three jointly participate in the PDU session establishment procedure.

We mainly show the MEN-SMF and MEN-AMF of the control plane in Fig. \ref{SCLER-FIG1}. They are enhancements and improvements of SMF and AMF in the 5G core network. They mainly participate in the establishment of UE Multi-Access PDU sessions. MEN-AMF Responsible for handling UE connection, registration as well as routing session management messages from the UE to the appropriate MEN-SMF. MEN-SMF manages PDU sessions by sending the signaling to different 6G access points. UPF on the user plane carries workflow data and can be regarded as a Layer 3 router, which receives and forwards the data packets between the RAT and the DN. Untrusted non-3GPP access Wifi shall be connected to the 3GPP Core Network via a Non-3GPP InterWorking Function (N3IWF)\cite{3gpp.23.502}.
The 5G network defines a PDU session as a logical connection between the UE and the DN. A part of the PDU session is carried on the wired resources of the core network, and a part exists on the wireless resources. The performance of the PDU session is constrained by the physical bearer, but the transmission process itself ignores the resource form. The existing mobile network supports the UE to access different DNs to establish corresponding multiple PDU sessions, and can also access a single DN to establish multiple PDU sessions. However, the current 5G does not support the establishment of physical or geographic separation of multiple PDU sessions for a single workflow. In Fig. \ref{SCLER_fig3} We give the procedure of establishing a Multi-Access PDU Session, the appendix A has a detailed procedure. Multiple PDU sessions provide services for a single service and are connected to the same DN. UPF-A is an anchor UPF that receives data packets to and from DNs for UEs\cite{ha2019support}. The splitting, aggregation, encoding, and decoding of data on the DN side are processed in SCLER-UPF-A.

\begin{figure*}[!t]
	\centering
	\includegraphics[width=7 in]{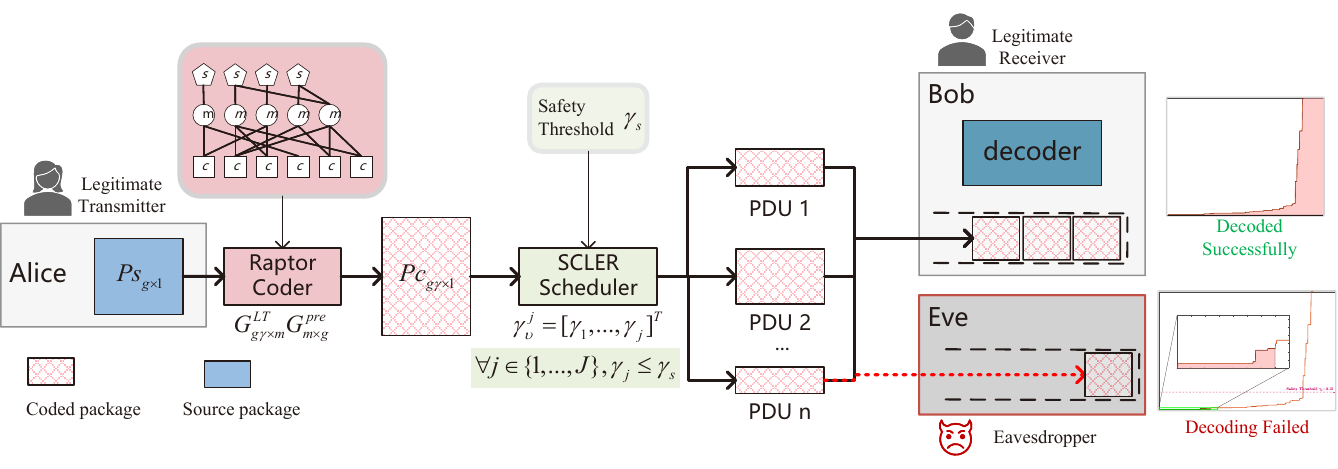}
	\caption{SCLER Scheme}
	\label{SCLER_Scheme}
\end{figure*}

\section{System Model}
As depicted in Fig.\ref{SCLER-FIG2}, SCLER creates three PDU sessions on the edge network to provide a downlink transmission strategy with low latency and security considerations from the DN to the UE. Service flows that randomly arrive at the DN's application layer are accumulated in the buffer and queued. The variable block length encoder selects $g$ data packets for encoding, and the encoded packets are distributed to three PDU sessions according to the k-out-of-N weight for transmission. The legitimate receiver receives the $Pc$ through the diversity interface and completes the decoding process. However, malicious network elements, eavesdroppers, and other unauthorized entities only receive part of the information and cannot complete the decoding.

The SCLER strategy divides time into time slots and delivers the DN's downlink service flow to the legal receiving terminal reliably with low delay while also considering security aspects. This section discusses the system model from three aspects: the modeling of the transmission strategy, the modeling of the transmission principle, and the key measurement indicators of the system.
\subsection{Encoding Transmission Model}
Under the SCLER strategy, we conduct Raptor encoding on a generation of $g$ data packets in each time slot\footnote{assuming that all data packets have the same length and carry $\phi $ bit information.}.  As shown in Fig.\ref{SCLER_Scheme}, by leveraging the fountain code prefix property, the redundancy of the encoder output can be adjusted by $T$. Assuming the encoder produces $N$ data packets, the $j$th-PDU session selects $k$ packets from them for transmission. We define the scheduling weight coefficient as ${\gamma _j} = k/N$, and the scheduling weight vector of SCLER as $\gamma _\upsilon ^j = {[{\gamma _1},...,{\gamma _j}]^T}$.
The SCLER strategy involves the selection of $g$ data packets to be encoded in time slot $T$, and then the distributing the data using a scheduling strategy $\gamma _\upsilon ^j$. 

In this study, we represent the SCLER action for each time slot by a two-tuple $(g,\gamma _\upsilon ^j)$, $i.e.$, $(g[n] = g,\gamma _\upsilon ^j[n] = \gamma _\upsilon ^j)$, indicates that the SCLER action in the $n$-th time slot is $g,\gamma _\upsilon ^j$. Assuming that the number of data packets in each generation is bounded by $B$, that is, $g[t] \in  \{0,1,...,B\} $. The weight value ${\gamma _j}[n]$ for a single session changes in step sizes of ${(\varsigma T)^{ - 1}}$, where $\varsigma$ is the encoding rate coefficient that indicates the number of codes generated per unit time, 
${\gamma _j} \in \{ \frac{1}{{\varsigma T}},\frac{2}{{\varsigma T}},...,1\}$, $0 < {\gamma _j} \le 1$.

We define the value space of the weight vector $\gamma _\upsilon ^j$ as $\Gamma _\upsilon ^j$, such that $\gamma _\upsilon ^j \in \Gamma _\upsilon ^j$. The space size is ${N_j} = {(\varsigma T)^j}$, and $\gamma _\upsilon$ is obtained by $ \odot _{i = 1}^{{N_L}}\gamma _\upsilon ^j({\gamma _i})$ here, $ \odot $ denotes the Cartesian product. We have $1 \le {\gamma _\upsilon } < {N_L}$, where ${N_L}$ represents the number of PDU sessions established by SCLER. To ensure a sufficient breadth of PDU session selection, we require ${N_L} \ge 2$. Each PDU session described in SCLER can be modeled as a packet erasure channel. Let $TX[n]$ be the number of packets sent by the transmitting end and $RX[n]$ be the number of packets received by the receiving end
\begin{equation}\label{PDU-channel-model}
RX[n] = \varepsilon [n]TX[n].
\end{equation}
We assume that $TX[n]$, has a probability of $1 - {P_\varepsilon }$ being successfully delivered to $RX[n]$. Otherwise, the packet will be erased. This erasure can be due to various factors, such as subsequent datagrams being discarded due to high router load at the network layer, non-confirmed retransmission triggered by a new PDU session timer after mobile reconnection, transmission errors detected by the bottom layer such as the data link layer, and discarded packets, etc\cite{julio2020r,yang2022network}. We denote $\varepsilon _\upsilon ^j = [{\varepsilon _1},...,{\varepsilon _j}]$ to represent the erasure probability of different paths, ${c_j}[n] = 0/1$ to indicate whether the $j$th PDU session is lost or successfully transmitted at time slot n, ${\mathbb{P}} \{ {c_j} = 0\}  = {\varepsilon _j}$, $c_j$ is independent and identically distributed in different time slots of the same PDU session.

\subsection{Threat Analysis and SCLER Encoding Principle}
In this subsection,  we first analyze the potential information leakage associated with the multiplex transmission method based on Raptor encoding. We then provide a description of the passive attackers that we aim to defend against in this paper.
Finally, we introduce the Raptor encoding transmission principle employed by SCLAR.

\subsubsection{Threat Analysis}
In this subsection, we describe the attack capabilities of passive attackers \cite{boualouache2022federated}, who do not tamper with or discard data like active attackers, but instead copy, or store stolen user business data. Specifically, we consider a generalized eavesdropper who aims to obtain user data carried on the transmission path between the UE and the DN, without considering the specific communication layer of the eavesdropping behavior(gNB or Network Function), as shown in Fig.\ref{SCLER-FIG2}. By clarifying the attack capabilities of passive attackers, we can better understand the scope of our defense goals.

Taking a B5G edge network with $N_L$ disjoint PDU sessions as an example, where $N_j$ represents the number of nodes contained in the $j$th PDU session. If $M$ nodes are attacked, it may result in only some nodes being under attack, which is a subset of the ${N_L}$ PDU sessions. This will lead to some information leakage, and we study the probability of information leakage. Where
\begin{equation}\label{leakage-percent}
	\begin{split}
	{\mathbb{P}} [M] &\ = \sum _{{\varpi _1} = 0}^{min({N_1},M)} \cdots \sum _{{\varpi _{{N_L} - 1}} = 0}^{min({N_j} - 1,M - \sum _{i = 1}^{{N_j} - 2}{\varpi _i})}\\
	&\ \prod _{j = 1}^{{N_L}}(\begin{array}{*{20}{c}}
		{{N_j}}\\
		{{\varpi _j}}
	\end{array})p_j^{{\varpi _j}}{(1 - {p_j})^{({N_j} - {\varpi _j})}}u({n_{{N_L}}} - {\varpi _{{N_L}}})
	\end{split}
\end{equation}
Assuming that ${\varpi_j}$ nodes are attacked in the jth PDU session, ${\varpi _{{N_L}}} = M - \sum _{j = 1}^{{N_L} - 1}{\varpi _j}$, $u( \cdot )$ refers to
\begin{equation}
	u(x) = \begin{cases}
		1, & \text{if } x \geq 0 \\
		0, & \text{otherwise}
	\end{cases}
\end{equation}
 the probability of M nodes being attacked is given by 
\begin{equation}
	\begin{split}
{\mathbb{P}} &\ [{\varpi _1},{\varpi _2} \ldots ,{\varpi _{{N_L}}},M] = \\
&\ \prod _{j = 1}^{{N_L}}(\begin{array}{*{20}{c}}
	{{N_j}}\\
	{{\varpi _j}}
\end{array})p_j^{{\varpi _j}}{(1 - {p_j})^{({N_j} - {\varpi _j})}}u({n_{{N_L}}} - {\varpi _{{N_L}}}),
\end{split}
\end{equation}
where ${\varpi _{{N_L}}} = M - \sum _{i = 1}^{{N_L} - 1}{\varpi _i}$ , and the marginal probability ${\mathbb{P}} [M]$ is obtained by summarizing all possible values ${\varpi _i}$ and $i = 1,2,...,{N_L} - 1$.

For the subset ${\Upsilon _{{n_L}}}$ of multi-PDU sessions with $n_L$ paths, which contains ${l_1},{l_2},...,{l_{{n_L}}}$, the probability of $M$ nodes being attacked is:
\begin{equation}
	\begin{split}
	{\mathbb{P}}[{\Upsilon _{{n_L}}},&\ M] =  \sum\nolimits_{{\varpi _l}_{_1} = 0}^{\min ({N_{{l_1}}},M)}  \ldots  \sum\nolimits_{{\varpi _l}_{_{{n_L}} - 1} = 0}^{\min ({N_{{{_l}_{_{{n_L}} - 1}}}},M - \sum\nolimits_{l = {l_1}}^{{l_{_{{n_L}} - 2}}} {{\varpi _l}} )}\\
	&\ {\prod\nolimits_{j = {l_1}}^{{l_{{n_L}}}} {\left( {\begin{array}{*{20}{c}}
					{{N_j}}\\
					{{\varpi _j}}
			\end{array}} \right)p_j^{{\varpi _j}}{{\left( {1 - {p_j}} \right)}^{({N_j} - {\varpi _j})}}u({N_l}_{_{{n_L}}} - {\varpi _l}_{_{{n_L}}})} }\\
		&\ {\rm{ }}\prod\nolimits_{\Psi  = 1}^{{N_L}} {{{(1 - {p_\Psi })}^{{n_\Psi }}}} 
		\end{split}
\end{equation}

where ${\varpi _{{l_{{n_L}}}}} = M - \sum _{j = {l_1}}^{{l_{{n_L}}} - 1}{\varpi _j}$, and $\Psi  \ne {l_1},{l_2},...,{l_{{n_L}}}$
Therefore, the probability that only $M$ nodes in subset ${\Upsilon _{{n_L}}}$ in ${N_L}$ are attacked is $P[{\Upsilon _{{n_L}}}|{N_{node}} = M] = \frac{{P[{\Upsilon _{{n_L}}},{N_{node}} = M]}}{{P[{N_{node}} = M]}}$.

Suppose there are ${N_L}$ disjoint PDU sessions in the B5G edge computing network, denoted by the node set $L = \{{l_1},...,{l_j},...  ,{l_{{N_L}}} \} $ for this MEN.   Here, ${N_L} \in \{ 1,...  ,J\}$ indicates that at most $J$ PDU sessions can be established, as shown in Fig.\ref{SCLER-FIG1} is an example of establishing 3 disjoint PDU sessions \footnote{in Fig.\ref{SCLER-FIG1}, we set the UPF anchor with DN together, and UPF-A is not drawn.}. ${l_j}$ represents the set of nodes contained in a single PDU session, ${l_j}=\{n_1^j,n_2^j,.. .,n_i^j,...  ,n_{{N_j}}^j\} $, where ${N_j} \in \{ 1,...,I\}$, $I$ represents the upper limit of nodes contained in a single PDU session. We consider the eavesdropper ${E_\kappa }$, who can erode at most ${N_j}$ nodes on the single $j$th PDU session and cannot eavesdrop on nodes other than the current PDU session,$i.e.$, $\{ \{ L\} \backslash \{ {l_j}\} \} $, ${\Delta _\kappa } = \{ {\delta _1},{\delta _2},...,{\delta _n}\} $ represents the eroded node set\footnote{Including malicious network functions, and eavesdropping on gNB, this article regards the erosion of a single node.}, satisfying
\begin{equation}\label{attack-model}
	\begin{split}
		&\ \forall \Delta_\kappa \subseteq L, \{ \Delta_\kappa \cap l_j \} \neq \varnothing, \exists j \in \{ 1,...,J \} , \quad \\
	&\ \text{s.t.} \quad \Delta_\kappa \subseteq l_j, \text{ and } \Delta_\kappa \cap \{ L / l_j \} = \varnothing
	\end{split}	
\end{equation}
Furthermore, there is no interaction or cooperation between all eavesdroppers ${E_\kappa },\kappa \in \left( {1,K } \right]$,  ${E_\kappa }$ will not be assisted by any other eavesdropper.  This assumption is reasonable in mobile networks \cite{huang2017coding}.

%	$$\forall \kappa \in {1,2,\dots,K},\ \exists M_j \subseteq N, \text{ s.t. } E_j \subseteq M_j$$
%
%$$\forall E_j \in N, \exists M \in N, \exists B \subseteq M \cap N, B \neq \varnothing : \forall n \in B, n \in M \cap N$$
%
%$\forall {\Delta _\kappa } \subseteq L,{\rm{s}}{\rm{.t}}{\rm{. }}\exists j \in \{ 1,...,J\} {\Delta _\kappa } \subseteq {l_j},and{\Delta _\kappa } \cap \{ L/{l_j}\}  = \emptyset $

\subsubsection{SCLER Encoding Principle}

\begin{algorithm}[!t]
	\caption{Raptor Code Encoding Operation}
	\label{alg:algorithm1}
	\begin{algorithmic}[1]
		\REQUIRE Read $Ps = {(P{s_1},P{s_2},...,P{s_g})^T}$ from the queue, ${\Omega _d} = ({\Omega _1},{\Omega _2},...,{\Omega _{\max }})$, Waiting window $w$;
		\ENSURE {$\bar Pc$}
		\STATE $P{c_{g\gamma  \times 1}} \leftarrow {0_{g\gamma }}$;\
		\STATE $G_{g\gamma  \times m}^{Raptor} = G_{g\gamma  \times m}^{LT}G_{m \times g}^{pre}$;\
		\STATE $P{c_{g\gamma  \times 1}} = G_{g\gamma  \times m}^{Raptor}P{s_{g \times 1}}$;\
		\STATE $\bar Pc \leftarrow P{c_{g\gamma  \times 1}}$;\
		\IF{$w>0$}
		\STATE the transmission is completed for $w$ time slots;\
%		\STATE unresponsive mode, ;
		\FOR{$ACK \ne nil$}
		\STATE $P{c^{fix}} \leftarrow G_{1 \times g}^{fix}P{s_{g \times 1}}$\;
		\STATE $\bar Pc \leftarrow append[\bar Pc,P{c^{fix}}]$\;
		\ENDFOR
		\ELSE 
		\STATE no response mode;
		\ENDIF
%		\WHILE{condition}
%		\STATE Statement
%		\ENDWHILE
		\STATE \textbf{return} {$\bar Pc$};
	\end{algorithmic}
\end{algorithm}

As shown in Fig.\ref{SCLER_Scheme}, we designed an encoding transmission method based B5G for UE-DN server flow, which encodes $Ps$ arriving at the application layer and distributes $Pc$ to multiple PDU sessions. The method involves a Raptor\cite{shokrollahi2006raptor} encoder that performs two encoding processes on the selected $g$ $Ps$: an external encoder (pre-encoding) and an internal encoder (LT code\cite{luby2002lt}). Following the SCLER strategy, we propose Algorithm\ref{alg:algorithm1}, the method first selects $g$ source data packets($Ps$) to encode as one generation, at time slot $t$, the encoder of a generation of $g$ is denoted by ${\xi _t}(g,m,\Omega)$, we have $P{c_{g\gamma  \times 1}} = G_{g\gamma  \times m}^{LT}G_{m \times g}^{pre}P{s_{g \times 1}}$, The formula $G_{g\gamma  \times m}^{Raptor} = G_{g\gamma  \times m}^{LT}G_{m \times g}^{pre}$ defines the relationship between $Ps$ and $Pc$, where $G_{g\gamma  \times m}^{LT}$ and $G_{m \times g}^{pre}$ represent the LT encoding generator matrix and precoding generator matrix. (1) Precoding generates m intermediate symbols to ensure that all $Ps$ are covered, $G_{m \times w}^{pre} = {[{I_k}|{P_{m \times (k - w)}}]^T}$, where ${P_{m \times (k - w)}}$ is responsible for mixing $Ps$. (2)The internal encoding matrix $G_{g\gamma  \times m}^{LT}$ generates the final encoding symbol based on the degree distribution $\Omega (x)$, which represents the probability distribution of the number of $Ps$ selected each time, ${\Omega _d} = ({\Omega _1},{\Omega _2},...,{\Omega _{\max }})$, while $\Omega (x) = \Sigma _{d = 1}^{{d_{\max }}}{\Omega _d}{x^d}$. Encoder input $P{s_{g \times 1}} = {(P{s_1},P{s_2},...,P{s_g})^T}$, output $P{c_{g\gamma  \times 1}} = {(P{c_1},P{c_2},...,P{c_{g\gamma }})^T}$. 

The Maximum-Likelihood (ML) decoding algorithm is typically preferred for finite-length Raptor encoding over the Belief Propagation (BP) decoding algorithm \cite{ karp2004finite, shokrollahi2009theory}. The performance of ML decoding is comparable to that of Gaussian elimination \cite{wang2016performance}. To use ML decoding, a coefficient matrix is constructed with rows and columns corresponding to the $Pc$ and $Ps$. The coefficient matrix represents how the $Pc$ is linearly combined from the $Ps$. $g$ $Ps$ can be recovered from $g\gamma$ of $Pc$ if and only if ${(G_{g\gamma  \times m}^{LT}G_{m \times g}^{pre})_{g\gamma  \times g}}$ is full rank.
\begin{figure}[!t]
	\centering
	\begin{subfigure}[t]{0.45\columnwidth}
		\centering
		\includegraphics[width=\textwidth]{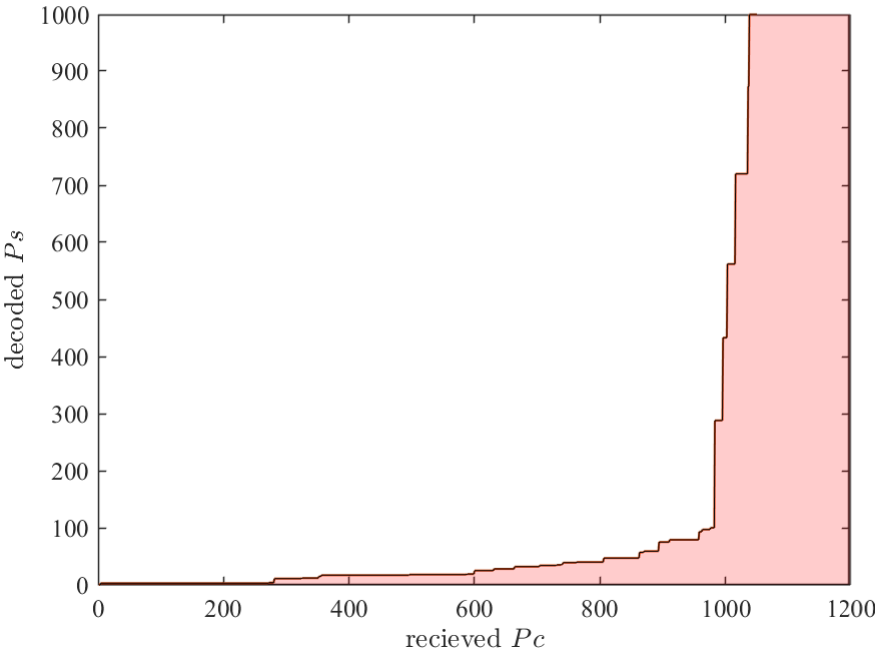}
		\caption{Legitimate Receiver.}
		\label{decoded-number:sub1}
	\end{subfigure}
	\hfill
	\begin{subfigure}[t]{0.45\columnwidth}
		\centering
		\includegraphics[width=\textwidth]{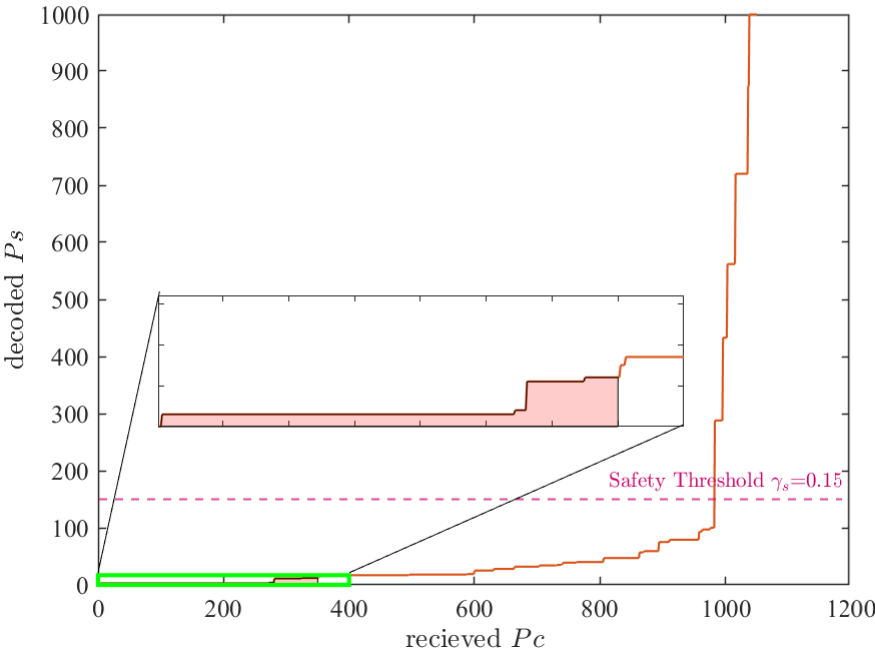}
		\caption{Eavesdropper.}
		\label{decoded-number:sub2}
	\end{subfigure}
	\caption{Number of $Ps$ decoded versus number of $Pc$ received for Receivers.}
	\label{decoded-number-versus}
\end{figure}

We conducted a MATLAB simulation of the RAPTOR codec and observed that the number of successfully decoded packets initially increases slowly with the number of received $Pc$, before reaching a sharp increase threshold\footnote{Note that our team has worked on improving the slope of the surge in the early stages\cite{yi2014achieving}, but we won't go into details in this article.}.      Once the threshold is reached, the entire block can be decoded quickly.   In other words, the number of successfully decoded $Ps$ is initially very low until the number of received coded packets reaches the threshold.   This principle can be used to defend against single-link eavesdropping attacks in CMT, we can limit the number of coded packets transmitted in each single PDU session to below this threshold.   This way, an attacker of a single PDU session cannot decipher the source data information effectively, ensuring the security of user data.

\subsection{Secure and Reliable Function and Queuing Model}
This subsection comprises two parts. In the first part, we analyze the reliability and security of a concurrent transmission system that employs Raptor coding. In the second part, we analyze a queuing coding model that corresponds to an eavesdropping scenario. For enumeration of all possible session state, we first employ a  ${2^{{N_L}}} \times {N_L}$ matrix $C$:
\begin{equation}\label{communication-matrix}
	C = {\left[ {\begin{array}{*{20}{c}}
				1&1& \ldots &0\\
				\vdots & \vdots & \ldots & \vdots \\
				1&0& \ldots &0
		\end{array}} \right]^{\rm{T}}}
\end{equation}
Each element ${c_{i,j}}$ in row $i$ and column $j$ represents the probability of successful ($c_{i,j} = 1$) or failed ($c_{i,j} = 0$) reception by the receiver through the $j$-th PDU session for the $i$-th transmission possibility. As the UE-DN workflow is transmitted through different PDU sessions, each individual message may experience varying degrees of delay and packet loss, which can be quantified using the latency-reliability-security function, 
\begin{equation}\label{latency-reliability-security-function}
	F_{SCLER}\left( {T,\gamma _\upsilon ^j,g} \right) = \sum\limits_{i = 1}^{{2^{{N_L}}}} {{g_i}} \prod\limits_{j = 1}^{{N_L}} {{H_j}} \left( {T,{\gamma _j}g} \right)
\end{equation}
%as illustrated in Eq. \ref{latency-reliability-security-function}, 
Based on Eq.\ref{latency-reliability-security-function}, the secure data transmission considering the existence of single-link eavesdroppers, $\forall j \in \{ 1,...,N_L\} $, ${\gamma _j} \le {\gamma _s}$, ${\gamma _s}$ is the threshold value for ensuring transmission security, and limits the upper bound of the weight assigned to each path to prevent eavesdroppers from obtaining sufficient private information. Where
\begin{equation}\label{equation-failout}
	{g_i} = \begin{cases}
		1,&{{\text{if}}\ \Sigma _{j = 1}^{{N_L}}{c_{i,j}}{\gamma _j}{\gamma _\tau } \ge {\gamma _d}} \;{{\rm{and}}\;\forall j \in \{ 1,...,{N_L}\}, {\gamma _j} \le {\gamma _s}} \\ 
		0,&{\text{otherwise.}} 
	\end{cases}
\end{equation}
${g_i}$ will exclude failed transmissions to ensure that only successfully decoded outputs are included. ${\gamma _d}$ is the threshold used to determine successful decoding.
${H_j}$ is defined as
\begin{equation}
	{H_j}\left( {T,{\gamma _j}g} \right) = \left\{ {\begin{array}{*{20}{c}}
			{{F_j}\left( {T,{\gamma _j}g} \right),}&{{\rm{if }}{c_{i,j}}{\rm{ = }}1}\\
			{1 - {F_j}\left( {T,{\gamma _j}g} \right),}&{{\rm{if }}{c_{i,j}}{\rm{ = }}0}
	\end{array}} \right.
\end{equation}
The definition of ${F_j}$ is based on the research by Erik \textit{et al.} \cite{strom20155g}. We delayed the transmission of PDU sessions to cut off the Gaussian distribution. The corresponding accumulated probability distribution, ${F_j}$, is shown in Figure \ref{result2:sub1}. ${F_j}\left({T, {\gamma_j}g}\right)$ means that at time T, the data packet transmission of the $j$-th PDU session with size ${\gamma_j}g$ can achieve a certain reliability. The extreme value of ${F_j}$ approaches ${1-P_e^j}$.

There is a correspondence between static latency and reliability (i.e., probability of successful data delivery).

By utilizing function Eq.\ref{latency-reliability-security-function}, both reliability and security can be guaranteed.
$\Lambda[t]$ denotes the number of $Ps$ arriving in the $t$-th timeslot. We assume that the application layer input $\{ \Lambda[t],t \ge 0\} $, the number of packets arriving in each slot is $i.i.d.$, the upper bound of $\Lambda[t]$ is $N$, denote the probability distribution of $\Lambda[t]$ as $\boldsymbol{\lambda}={[{\lambda _0},{\lambda _1},...,{\lambda _N}]^T}$, where we define ${\lambda _n} = {\mathbb{P}} \{ \Lambda[t] = n\} $, the average arrival rate as $\sum\nolimits_{n = 0}^{{N_\Lambda }} {n{\lambda _n}} $. Using a buffer with a size of $Z$, randomly arriving packets are accumulated, and $g$ packets are selected from the cache once the encoding of the previous block of packets is completed. Therefore, $q[t] \in \{ 0,1,...,{\rm Z}\} $, in the $[t+1]$-th timeslot, $q[t]$ evolves as\footnote{we define ${(x)^ * } = \max \{ x,0\} $} 
\begin{equation}\label{queue-enolves}
q[t + 1] = \min \{ {(q[t] - g[t])^ * } + \Lambda [t+1],{\rm Z}\}
\end{equation}
%%%%%%%%%%%%%%%%%%%%%%%%%%%%%%%%%%%%%%%%%%
\section{Trade-off Delay-Security For Variable Block-Length SCLER Strategy}

\subsection{Constrained Markov Decision Process Formulation}%问题模型
In the context of edge networks, ensuring both the security and low latency of data transmission is critical. To address these challenges, we propose a Raptor-encoded multi-session distributed delay-optimized transport method called SCLER that considers these constraints. To achieve this, we utilize the queue length as the system state information and propose a novel queue-aware variable block length encoding method that effectively models the SCLER strategy as a Constrained Markov Decision Process (CMDP). We formulate the problem as a constrained optimization problem to obtain the trade-off between delay and reliability, which is a crucial factor in designing efficient data transmission mechanisms in edge networks. Specifically, we introduce the SCLER strategies by leveraging the probabilities $f(q,g,\gamma _\upsilon ^j)$,

\begin{equation}\label{SCLER-strategies-probabilities}
	f(q,g,\gamma _\upsilon ^j) = {\mathbb{P}} \{ g[t] = g,\gamma _\upsilon ^j[t] = \gamma _\upsilon ^j|q[t] = q\}. 
\end{equation}
We assume a temporary steady-state condition where no PDU session is being established or released, that is $\forall q,\Sigma _{g = 1}^B{\Sigma _{\gamma _\upsilon ^j \in \Gamma _\upsilon ^{{N_j}}}}f(q,g,\gamma _\upsilon ^j) = 1$. Therefore, we can get the strategy function of SCLER, $\textbf{\emph{S}} = \left\{ {f\left( {q,g,\gamma _\upsilon ^j} \right):0 \le q \le M,0 \le g \le B,1 \le {\gamma _\upsilon } < {N_L},0 < {\gamma _j} \le 1} \right\}$, to prevent the sender buffer from overflowing or underflowing, $0 \le q[t] - g[t] \le M - {N_\Lambda }$, and  $\{ \gamma _\upsilon ^j \in \Gamma _\upsilon ^{{N_j}}|{{\mathbb{C}}_{\{ g > 0\} }} \le {\gamma _\upsilon } \le {N_L}{{\mathbb{C}}_{\{ g > 0\} }},0 < {\gamma _j} \le {{\mathbb{C}}_{\{ g > 0\} }}\} $. The state of the system q evolves based on the Eq.\ref{queue-enolves}. We denote the overall redundancy of a generation of data by ${\gamma _\upsilon } = \sum {_{j = 1}^{{N_L}}} {\gamma _j}$. Given a particular strategy and corresponding system constraints, if $T \to \infty $, then $\mathop {\lim }\limits_{T \to \infty } F\left( {T,\gamma _\upsilon ^j,g} \right) = L \le 1 - {P_e}$, L denotes the upper limit of information security transmission, and the reliability value of the s ystem will not exceed its theoretical performance. Using the delay-safety function, we can identify the feasible region
\begin{equation}
{\cal F}(\epsilon) = \left\{ {\forall \gamma _\upsilon ^j,g,F\left( {T,\gamma _\upsilon ^j,g} \right)|1 - \epsilon \le F\left( {{\gamma _\tau }T,\gamma _\upsilon ^j,g} \right) \le 1 - {P_e}} \right\}
\end{equation}
The fountain code transmission scheme in edge networks results in a long-tailed latency distribution. To address this issue, the SCLER algorithm is proposed to reduce the average transmission delay and achieve a balance between reliability and security. Based on Little's Law, $\mathop {\lim }\limits_{T \to \infty } \frac{1}{T}\sum\nolimits_{t = 1}^T {E\{ \frac{{q[t]}}{\lambda }\} } $, this can be used as a latency metric for MEN, which we describe by the probability of constraints violation ${{\mathbb{P}} vc}$:
\begin{equation}
{\mathbb{P}vc} = \mathop {\lim }\limits_{T \to \infty } \frac{1}{T}\sum\nolimits_{t = 1}^T {{\mathbb{P}} \{ \frac{{q[t]}}{\lambda } \ge {d^{{\rm{th}}}}\} } 
\end{equation}
where ${d^{{\rm{th}}}}$ is the threshold of the latency indicator, ${d^{{\rm{th}}}} \in \left[ {0,\infty } \right)$
Building upon the SCLER strategy described above, this paper investigates the provision of low-latency, secure, and reliable transmission services. A comprehensive analysis is conducted, considering indicators such as information leakage rate, PDU session quality, occupied bandwidth, decoding function, and delay. The delay and security of the MEN in the B5G era are key performance indicators that are determined by the state $(q[n],\varepsilon _\upsilon ^j[n])$ and action $(g,\gamma _\upsilon ^j)$. Considering the limitations of service characteristics, this section proposes a constrained optimization problem based on CMDP, and the next section will analyze the trade-off between reliability and delay while considering security.

\subsection{The Optimal Delay-Reliability Trade-off under SCLER}%问题模型
In this subsection, we will focus on achieving a trade-off between transmission reliability and delay while ensuring data security in the face of Eq.\ref{attack-model}. We refer to Eq.\ref{latency-reliability-security-function} as the reliability function, and successful transmission of data in SCLER involves two parts: receiving and recovering. As shown in the Fig.\ref{SCLER-FIG2}, SCLER determines the transmission strategy $\{g,\gamma _\upsilon ^j\}$ based on the state information of the queue length $q[t]$. We describe the SCLER strategy $\textbf{\emph{S}}$ as a Markov decision process with the buffer queue $q(t)$, we define the transition probability as ${\lambda _{q,q'}} = {\mathbb{P}} \{ q[t + 1] = q'|q[t] = q\} $. Based on Eq.\ref{queue-enolves}, we formulate ${\lambda _{q,q'}}$ as follows:
\begin{equation}
	{\lambda _{q,q'}} = \sum\limits_{n = 0}^N {{\lambda _n}\sum\limits_{g = 0}^B {f(q,g,\gamma _\upsilon ^j)} } {{\mathbb{C}}_{\{ q' = \min \{ {{(q - g)}^*} + n,\;{\rm Z}\} }}\footnote{${{\mathbb{C}}_{\{  \cdot \} }}$ denotes the characteristic function.}
\end{equation}
The steady state distribution of the system state $q[t]$ of SCLER is formulated as follows: 
\begin{equation}\label{qu-state-dis}
	\left\{ {\begin{array}{*{20}{l}}
		 	\sum\nolimits_{q = \max \{ 0, q' - N\} }^{\min \{ {\rm Z}, q' + B\} } {\pi (q){\lambda _{q,q'}} = } \pi (q'), &\forall q' \in {\rm{\{ 0,1,}}...{\rm{,M\} }},\\
			\sum\nolimits_{q = 0}^M {\pi (q) = } 1. &  
	\end{array}} \right.
\end{equation}
The queue length, denoted as $q$ and $q'$, takes on values in the set $\{0, 1, \dots, {\rm Z}\}$, subject to the constraint that the summation of $\pi(q)$ over all possible values of $q$ is equal to one. The queue length is influenced by the random arrival of packets and selection of. To express the balance equations in Eq.\ref{qu-state-dis} as a matrix form, we represent the steady-state probability distribution $\bm{{\pi}}$ as a vector ${[{\pi _0},...,{\pi _{\rm Z}}]^T}$,
\begin{equation}
	\left\{ {\begin{array}{*{20}{l}}
			\Lambda_\textbf{\emph{S}} {\bm{{\pi}}} = {\bm{{\pi}}}, &\\
			 \bm{1}^{T}{\bm{{\pi}}} = 1, &
	\end{array}} \right.
\end{equation}
The matrix $\Lambda_\textbf{\emph{S}}$ is defined with the element ${\lambda _{q,q'}}$ at the $q$th row and $q'$th column. Additionally, the vector $\bm{1}$ is defined as a vector with all its components being 1. Based on the stationary distribution $\bm{{\pi}}$, the SCLER strategy's average delay can be expressed using Little's law. Specifically, the average delay ${L_\textbf{\emph{S}}}$ based on the SCLER strategy $\emph{S}$ is
\begin{equation}
	{\bar {L}_\textbf{\emph{S}}} = \frac{1}{\lambda }\sum\limits_{q = 0}^{\rm Z} {q{\bm{{\pi}} _q}} 
\end{equation}
The occupied PDU session resource occupancy is expressed as
\begin{equation}
	{W_\textbf{\emph{S}}} = \sum\limits_{q = 0}^{\rm Z} {\sum\limits_{g = 0}^B {W_\textbf{\emph{S}}^j(g,\gamma _\upsilon ^j){f_{q,g,\gamma _\upsilon ^j}}} } {\pi _q}
\end{equation}

SCLER presents an optimal trade-off scheme for delay-reliability that takes into account security considerations, based on the average delay and bandwidth occupancy. The proposed approach formulates an optimization problem of $Ps$ encoding and CMT, with the queue length serving as state information, and subject to reliability and security constraints.

\begin{subequations}\label{delayaimMC}\small
	\begin{align}
		\mathop {\min }\limits_{\{ {f_{q,g,\gamma _\upsilon ^j}},{\pi _q}\}   }\;\; 
		&{\bar {L}_\textbf{\emph{S}}} \label{delayaimMC:1A} \\
		{{\rm{s}}{\rm{.t}}{\rm{.}}}\quad\quad 
		& {F_\textbf{\emph{S}}} \ge r^{{\rm{th}}}\label{delayaimMC:1B}\\
		& {W_\textbf{\emph{S}}}\le W_j^{\rm{th}} \label{delayaimMC:1C}\\
		& \Lambda_\textbf{\emph{S}} {\bm{{\pi}}} = {\bm{{\pi}}} \label{delayaimMC:1D}\\
		& \bm{1}^{T}{\bm{{\pi}}} = 1\label{delayaimMC:1e}\\
		& \sum\limits_{g = 0}^B {\sum\limits_{\gamma _\upsilon ^j \in \Gamma _\upsilon ^{{N_j}}} {{f_{q,g,\gamma _\upsilon ^j}}} }  = 1,\qquad \forall q\label{delayaimMC:1F}\\
		& {\pi _q} \ge 0, {{f_{q,g,\gamma _\upsilon ^j}}}\ge 0,\qquad\forall q,g.\label{delayaimMC:1G}
	\end{align}
\end{subequations}

The goal of SCLER is to find the optimal delay under the constraint of bandwidth $W_j^{\rm{th}}$ in the MEN. In order to achieve the optimal trade off between delay and reliability with secure consideration, we have transformed the encoding CMT problem into an Linear Programming (LP) problem:
\begin{subequations}\label{delayaimMCmatrix}\small
	\begin{align}
		\mathop {\min }\limits_{\{ {x_{q,g,\gamma _\upsilon ^j}} \}   }\quad 
		&\frac{1}{\lambda }\sum\limits_{q = 0}^{\rm{Z}} {\sum\limits_{g = 0}^B {\sum\limits_{\gamma _\upsilon ^j \in \Gamma _\upsilon ^{{N_j}}} q{{x_{q,g,\gamma _\upsilon ^j}}} } }  \label{delayaimMCmatrix:1A} \\
		{{\rm{s}}{\rm{.t}}{\rm{.}}}\quad\quad 
		& \sum\limits_{q = 0}^M {\sum\limits_{g = 0}^B {\sum\limits_{\gamma _\upsilon ^j \in \Gamma _\upsilon ^{{N_j}}} {F(g,\gamma _\upsilon ^j){x_{q,g,\gamma _\upsilon ^j}}} } }  \ge {r^{\rm{th}}}\label{delayaimMCmatrix:1B}\\
		& \sum\limits_{q = 0}^{\rm{Z}} {\sum\limits_{g = 0}^B {\sum\limits_{\gamma _\upsilon ^j \in \Gamma _\upsilon ^{{N_j}}} {W_\textbf{\emph{S}}^j(g,\gamma _\upsilon ^j){x_{q,g,\gamma _\upsilon ^j}}} } }  \le W_j^{{\rm{th}}} \label{delayaimMCmatrix:1C}\\
		& \sum\nolimits_{q = 0}^{\rm Z} \sum\limits_{n = 0}^N {\sum\limits_{g = 0}^B {\sum\limits_{\gamma _\upsilon ^j \in \Gamma _\upsilon ^{{N_j}}} {{\lambda _n}{x_{q,g,\gamma _\upsilon ^j}}} } } {{\mathbb{C}}_{\{ q' = \min \{ {{(q - g)}^*} + n,\;{\rm{Z}}\} }} \nonumber\\
		&  = \sum\limits_{g = 0}^B {\sum\limits_{\gamma _\upsilon ^j \in \Gamma _\upsilon ^{{N_j}}} {{x_{q',g,\gamma _\upsilon ^j}}} }    \label{delayaimMCmatrix:1D}\\
		& \sum\limits_{q = 0}^{\rm Z} {\sum\limits_{g = 0}^B {\sum\limits_{\gamma _\upsilon ^j \in \Gamma _\upsilon ^{{N_j}}} {{x_{q,g,\gamma _\upsilon ^j}}} }  = 1} \label{delayaimMCmatrix:1E}\\
		& {x_{q,g,\gamma _\upsilon ^j}}\ge 0,\qquad \forall q,g
		\label{delayaimMCmatrix:1F}\\
		& {x_{q,g,\gamma _\upsilon ^j}}=0,\qquad\forall q - g < 0\;or\;q - g > ({\rm Z} - N)
		\label{delayaimMCmatrix:1G}
	\end{align}
\end{subequations}
we define the optimization variable ${x_{q,g,\gamma _\upsilon ^j}}$ as
${\pi _q}{f_{q,g,\gamma _\upsilon ^j}}$, its long-run average can be expressed as $\mathop {\lim }\limits_{T \to \infty } {\textstyle{1 \over T}}\sum\nolimits_{t = 1}^T {\mathbb{P} \{ q[t] = q,g[t] = g,\gamma _\upsilon ^j = \gamma _\upsilon ^j[t]\} } $ , one can obtain the optimal trade-off between delay and reliability for SCLER by solving the LP problem Eq.\ref{delayaimMCmatrix} through the variable ${x_{q,g,\gamma _\upsilon ^j}}$. 

In algorithm \ref{alg:algorithm2}, we transform the constrained optimization problem Eq.\ref{delayaimMCmatrix} into a matrix form, we set the combination of solution space $\{ {x_{q,g,\gamma _\upsilon ^j}} \}$ as a vector $\vec x$, 
%, and reformulate Eq.\ref{delayaimMCmatrix} as
%\begin{equation}
%	\mathop {\min }\limits_{\vec x \in \{ {x_{q,g,\gamma _\upsilon ^j}}\} } \{ \overrightarrow L _S^T\vec x\;|\;\overrightarrow W _S^T\vec x \le {W^{{\rm{th}}}},\overrightarrow F _S^T\vec x \ge {r^{{\rm{th}}}}\} 
%\end{equation}
the feasible region in matrix form as
\begin{equation}
	{\Re } = \left\{ \begin{array}{l}
		\vec x \;|\;\textbf1_{\left| \vec x \right|}^T\vec x = 1,\;{\textbf{\emph{H}}_\textbf{\emph{S}}}\vec x = {\textbf0_{M + 1}},\\
		0 \le {x_{q,g,\gamma _\upsilon ^j}} \le {{\mathbb{C}}_{\{ g \in {\rm{{\cal N}}}(q),\ {\gamma _\upsilon ^j \in \Gamma _\upsilon ^{{N_j}}}\} }}
	\end{array} \right\}.\footnote{${\textbf1_N}$ and ${\textbf0_N}$ denote $N$ dimensional all-ones and all-zeros vectors, respectively.}
\end{equation}
Eq.\ref{delayaimMCmatrix:1D} are also presented in the form of coefficient matrix ${\textbf{\emph{H}}_\textbf{\emph{S}}}$, where ${\textbf{\emph{H}}_\textbf{\emph{S}}} = {\dot{\textbf{\emph{H}}}_\textbf{\emph{S}}} - {\ddot{\textbf{\emph{H}}}_\textbf{\emph{S}}}$, let ${\dot{\textbf{\emph{H}}}_\textbf{\emph{S}}}$ and ${\ddot{\textbf{\emph{H}}}_\textbf{\emph{S}}}$ denote the left-hand and right-hand coefficients of Eq. (\ref{delayaimMCmatrix:1D}), respectively. We utilize  $\textbf{\emph{S}}$ to obtain the optimal trade-off solution $x_{q,g,\gamma _\upsilon ^j}^ \wedge $ between delay and reliability, where $x_{q,g,\gamma _\upsilon ^j}^ \wedge $ is obtained through ${\pi _q}f_{q,g,\gamma _\upsilon ^j}^ \wedge $. Thus, the optimal strategy ${S^ \wedge }$ can be expressed as ${S^ \wedge } = \{ f_{q,g,\gamma _\upsilon ^j}^ \wedge :\forall q,g,\gamma _\upsilon ^j\} $, where $f_{q,g,\gamma _\upsilon ^j}^ \wedge $ is defined as:
\begin{equation}\label{optimal-function}
	{f_{q,g,\gamma _\upsilon ^j}^ \wedge} = \begin{cases}
		\frac{{x_{q,g,\gamma _\upsilon ^j}^ \wedge }}{\bm{\pi}_q^\wedge} &{{\text{if}}\ {\bm{\pi}_q^\wedge}>0} \\ 
		\mathbb{C}_\{ g = g_q^{\max }\}  &{\text{if}\ {\bm{\pi}_q^\wedge}=0},
	\end{cases}
\end{equation}
As a consequence, the optimal delay-reliability trade off under secure consideration for SCLER is obtained.

\begin{algorithm}[!t]
	\caption{Algorithm to generate constraints matrix $\textbf{\emph{H}}_\textbf{\emph{S}}$}
	\label{alg:algorithm2}
	\begin{algorithmic}[1]
		\REQUIRE $\boldsymbol{\lambda}={[{\lambda _0},{\lambda _1},...,{\lambda _N}]^T}$;
		\ENSURE {$\textbf{\emph{H}}_\textbf{\emph{S}}$}
		\FOR{$q = 0$ to $M$}
		\STATE $\dot {\textbf{\emph{H}}}_\textbf{\emph{S}}^q \leftarrow {0_{M + 1}}0_{B + 1}^T$, $\ddot {\textbf{\emph{H}}}_\textbf{\emph{S}}^q \leftarrow {0_{B + 1}}$;			
		\FOR {$g = 0$ to $B$}
		\STATE$\dot {\textbf{\emph{H}}}_\textbf{\emph{S}}^{q,g} \leftarrow {0_{\max \left\{ {q - g + N,M} \right\} + 1}}$;
		\STATE$\dot {\textbf{\emph{H}}}_\textbf{\emph{S}}^{q,g}(q - g + 1:q - g + N + 1) \leftarrow \boldsymbol{\lambda} $;
		\STATE$\dot {\textbf{\emph{H}}}_\textbf{\emph{S}}^q(1:M,g + 1) \leftarrow {{\mathbb{C}}_{\{ g \in {\rm{{\cal N}}}(q)\} }}\ddot {\textbf{\emph{H}}}_\textbf{\emph{S}}^{q,g}(1:M)$;
		\STATE$\ddot {\textbf{\emph{H}}}_\textbf{\emph{S}}^q(g + 1) \leftarrow {{\mathbb{C}}_{\{ g \in {\rm{{\cal N}}}(q)\} }}$;
		\ENDFOR
		\STATE$\dot {\textbf{\emph{H}}}_\textbf{\emph{S}}^q \leftarrow \dot {\textbf{\emph{H}}}_\textbf{\emph{S}}^q$, $\ddot {\textbf{\emph{H}}}_\textbf{\emph{S}}^q \leftarrow {e_{M + 1,q + 1}}{(\ddot {\textbf{\emph{H}}}_\textbf{\emph{S}}^q)^T}$;
		\STATE${\dot {\textbf{\emph{H}}}_\textbf{\emph{S}}} \leftarrow [{\dot {\textbf{\emph{H}}}_\textbf{\emph{S}}},\dot {\textbf{\emph{H}}}_\textbf{\emph{S}}^q]$, ${\ddot {\textbf{\emph{H}}}_\textbf{\emph{S}}} \leftarrow [{\ddot{\textbf{\emph{H}}}_\textbf{\emph{S}}},\ddot {\textbf{\emph{H}}}_\textbf{\emph{S}}^q]$;
		\ENDFOR
		\STATE ${\textbf{\emph{H}}_\textbf{\emph{S}}} \leftarrow {\dot{\textbf{\emph{H}}}_\textbf{\emph{S}}} - {\ddot{\textbf{\emph{H}}}_\textbf{\emph{S}}}$;
		%		\STATE \textbf{return} {$\bar Pc$};
	\end{algorithmic}
\end{algorithm}

%%%%%%%%%%%%%%%%%%%%%%%%%%%%%%%%%%%%%%%%%%
\subsection{The threshold-based Strategy}
We aim to offer edge network users a secure, high-reliability, and low-latency service. To achieve this, we propose transforming the LP problem into a reliability-delay pairing based on the reliability function (Eq.\ref{latency-reliability-security-function}). As a result, the LP problem can be redefined as:
\begin{subequations}\label{delayaimpair}\small
	\begin{align}
		\mathop {\min }\limits_{(\hat r,\hat L) \in \aleph  }\;\; &{\hat L} \label{delayaimpair:1A} \\
		{{\rm{s}}{\rm{.t}}{\rm{.}}}\quad\quad 
		& {\hat r} \ge r_j^{{\rm{th}}}\label{delayaimpair:1B}\\
		& {W_\textbf{\emph{S}}}\le W_j^{\rm{th}} \label{delayaimpair:2C}
	\end{align}
\end{subequations}
the set $\aleph $ comprises all admissible "delay-reliability" pairs under the SCLER strategy $\textbf{\emph{S}}$. Specifically, $\aleph $ can be obtained on the two-dimensional plane of delay and reliability, where delay and reliability are determined by Eq.\ref{delayaimMCmatrix:1A} and Eq.\ref{delayaimMCmatrix:1B}, respectively. Each point in $\aleph $ corresponds to a particular type of SCLER strategy $\textbf{\emph{S}}$ subject to the constraints specified by Eqs.\ref{delayaimMCmatrix:1C}-\ref{delayaimMCmatrix:1G}.

Using variable block length coding under given PDU session conditions $W_j^{\rm{th}}$ and arbitrary initial queue conditions, we obtain an optimal delay–reliability trade-off $f_{q,g,\gamma _\upsilon ^j}^ \wedge$. We also examine how our SCLER behaves under this optimal coding strategy ${S^ \wedge }$ and show that each queue length selects either maximum or minimum block lengths ($B_{g[t]}^{\max }$, $B_{g[t]}^{\min }$): maximum lengths improve coding efficiency and reliability Eq.\ref{latency-reliability-security-function}; packets dispersion CMT with a security threshold $ \gamma _s$ impose a lower limit on minimum $Pc$ encoding block lengths. When $g$ falls below this threshold $B_{g[t]}^{\min }$, we fill out coding blocks with randomly generated packets.
\begin{equation}
	\left\{ {\begin{array}{*{20}{l}}
			{f_{q,g,\gamma _\upsilon ^j}^ \wedge} \ge 0, &{\text{if}\ g_q^{\max }\ {\text{or}}\ g_q^{\min }}\\
			{f_{q,g,\gamma _\upsilon ^j}^ \wedge} = 0, & \text{otherwise}
	\end{array}} \right.
\end{equation}
The optimal SCLER encoding strategy exhibits characteristics of action-limited decision-making, where the strategy of S is constrained by state information and a threshold ${q^ \sim}$ related to $q[t]$: 
\begin{equation}
	\left\{ {\begin{array}{*{20}{l}}
			{f_{q,g_q^{\max },\gamma _\upsilon ^j}^ \wedge} = 1 &{\text{if}\ q>{q^ \sim }}\\
			{f_{q,g_q^{\min },\gamma _\upsilon ^j}^ \wedge} = 1 &{\text{if}\ q<{q^ \sim }}\\
			{f_{q,g_q^{\min },\gamma _\upsilon ^j}^ \wedge}+{f_{q,g_q^{\max },\gamma _\upsilon ^j}^ \wedge}=1 &{\text{if}\ q={q^ \sim }}
	\end{array}} \right.
\end{equation}
In addition, to achieve the optimal trade-off between delay and reliability, the allocation coefficient $\gamma _\upsilon ^j$ selects the maximum value within the allowable range of the safety threshold. As a result, we can obtain the minimum ${L_\textbf{\emph{S}}}$ while satisfying the reliability and safety constraint. The trade-off between delay and reliability can be obtained by adjusting the reliability constraints.

%%%%%%%%%%%%%%%%%%%%%%%%%%%%%%%%%%%%%%%%%%
\section{Performance Results}
We constructed a set of edge networks to evaluate our proposed SCLER.  The entire mobile core network is implemented based on the modified Free5GC.  The control plane and user plane network elements of the core network are deployed on two separate virtual machines running Ubuntu18.04 with 16GB memory.  We used two access points, two customized base stations (NR01), and one TP-LINK AX3000 Wi-Fi device.  We have modified the existing PDU session establishment process (see Appendix A) by transforming the core network control plane network element to support the establishment of multiple PDU sessions.  Our experimental section is divided into two parts: testing the basic environment and presenting the numerical results of the SCLER strategy.
\begin{figure}[h]
	\centering
	\begin{subfigure}[t]{0.47\columnwidth}
		\centering
		\includegraphics[width=\textwidth]{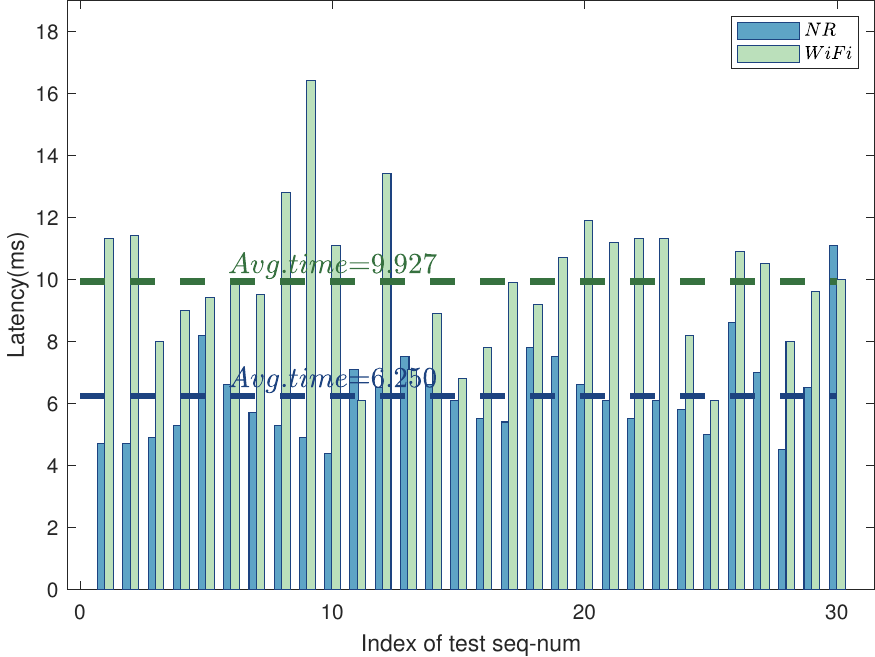}
		\caption{WiFi \& NR latency result.}
		\label{result1:sub1}
	\end{subfigure}
	\hfill
	\begin{subfigure}[t]{0.47\columnwidth}
		\centering
		\includegraphics[width=\textwidth]{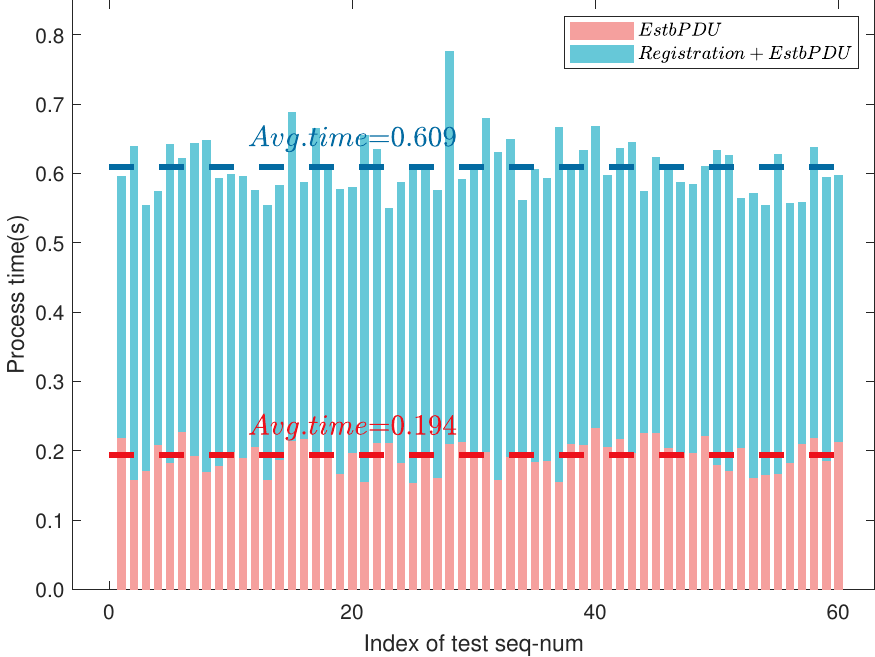}
		\caption{UE initialization procedure delay result.}
		\label{result1:sub2}
	\end{subfigure}
	\caption{DIEN performance test result.}
	\label{result1}
\end{figure}
\subsection{Basic Performance Results}
Fig.\ref{result1} shows the transmission performance test results of DIEN basic environment. As shown in Fig.\ref{result1:sub1}, we tested the transmission delay of NR and WiFi by sending an ICMP packet to the DN end on the UE. To determine the delay of the basic propagation of the PDU session that is not related to the carrying data packet. We recorded the delay of a single NR/WiFi access PDU session. The average delay of NR was 6.25ms, while that of WiFi was 9.927ms for 60 tests.

Furthermore, we conducted UE initialization procedure test Fig.\ref{result1:sub2} to measure the system's initialization delay. The experimental object is the procedure proposed by our Appendix A. 1.Registration procedure is usually triggered during power-on or switching flight mode. 2.Establishment of PDU session procedure triggers when the UE moves between gNB. 
We recorded the delay of the initialization process 30 times. The average time taken for Procedure1+2 and Procedure2 were 609ms and 194ms, respectively. The analysis of SCLER performance under the steady state after connecting the establishment of connection does not involve two initialization procedures.
\begin{figure}[h]
	\centering
	\begin{subfigure}[t]{0.47\columnwidth}
		\centering
		\includegraphics[width=\textwidth]{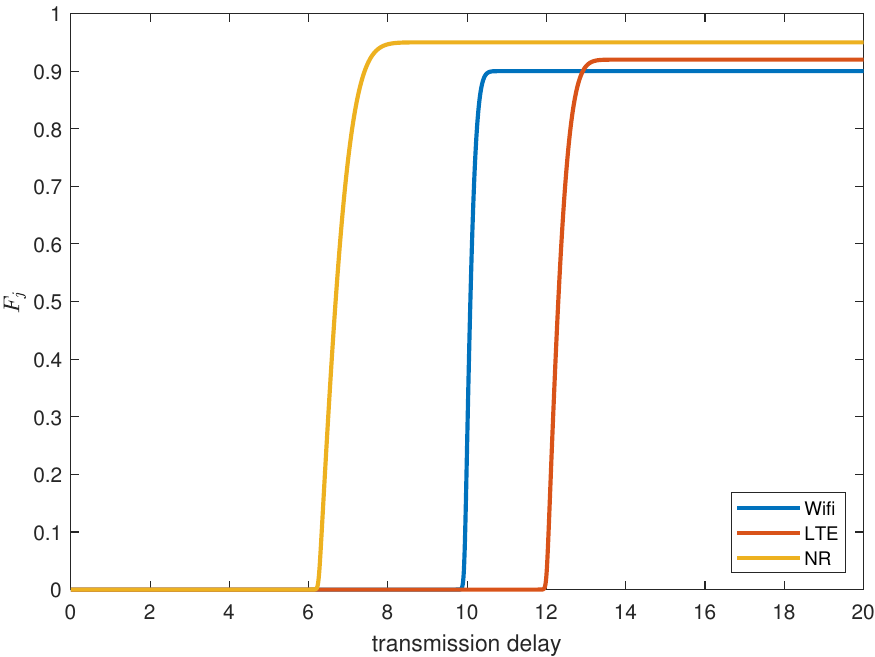}
		\caption{Latency-reliability curves ${F_j}$ for considered technologies for $\phi $ = 1500.}
		\label{result2:sub1}
	\end{subfigure}
	\hfill
	\begin{subfigure}[t]{0.47\columnwidth}
		\centering
		\includegraphics[width=\textwidth]{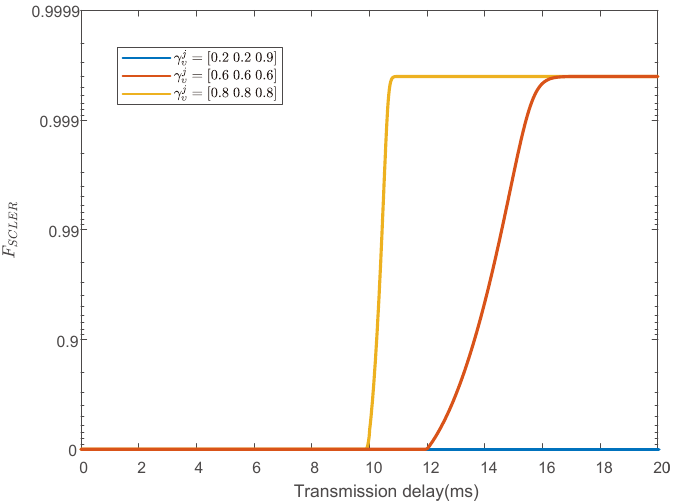}
		\caption{Delayed reliability function $F_{SCLER}$ with security consideration described in Eq.\ref{latency-reliability-security-function} .}
		\label{result2:sub2}
	\end{subfigure}
	\caption{Example showing the reliability diagram of a single PDU session and DIEN-based reliability function of the entire SCLER system.}
	\label{result2}
\end{figure}

As shown in Fig.\ref{result2}, we want to use ${F_j}$ to describe the delay characteristics of PDU sessions with different interfaces. The end-to-end transmission delay $T$ of the PDU session we consider may be affected by high load packet loss at the network layer\cite{wang2019event} and data link layer packet loss\cite{khatua2020application}. Therefore, without considering retransmission, $T$ should obey the Gaussian distribution, consequently we have $T=t_p+t_c$, where $t_c$ represents the data transmission delay, related to packet size and bandwidth, $i.e.\;{t_c} = W*{\phi ^{ - 1}}*1.25*{10^6}$, the propagation delay $t_p$ caused by protocol flow and physical signal propagation, corresponds to the position of the first non-zero point in Fig.\ref{result2:sub1}, where $t_p=[9.927, 12, 6.25]$ms, $W_j=[500,100,200]$Mbit/s, We regard the cumulative distribution of the delay probability as ${F_j}$. This indicates that the transmission reliability of the current session changes with the transmission delay, and its extreme value approaches $\{1-P_e\}$, where $P_e^j=[0.9, 0.92, 0.95]$, corresponding to the highest reliability that can be guaranteed. The rising rate of ${F_j}$ is related to $t_c$.

\begin{table*}[!t]
	\begin{center}
		\caption{Simulation Parameter Settings.}
		\label{tab2}
		\begin{tabular}{| c | c | c | c | c | c | c |}
			\hline
			\multicolumn{2}{c}{Packet Arrival Rate} & \multicolumn{2}{c}{Channel Transmission State} & \multicolumn{3}{c}{Packet Arrival Statistics} \\
			\hline
			${\mathbb{E}[\Lambda]}$ & $\boldsymbol{\lambda}={[{\lambda _{25}},...,{\lambda _{100}}]}$ & $r^{{\rm{th}}}$ & $\varepsilon _\upsilon ^j(\%)$ & ${\mathbb{E}[\Lambda]}$ & $\sigma^2$ & $\boldsymbol{\lambda}={[{\lambda _{25}},...,{\lambda _{100}}]}$ \\
			\hline
			25 & $[0.24,0.10,0.12,0.05]$ & $0.99$ & $[22, 25, 25]$ & 45 & 0.78 & $[0.39,0.42,0.11,0.06]$ \\
			\hline
			45 & $[0.34,0.27,0.16,0.11]$ & $0.999$ & $[9, 22, 10]$ & 45 & 1.16 & $[0.39,0.31,0.13,0.10]$ \\
			\hline
			60 & $[0.16,0.30,0.20,0.26]$ & $0.9996$ & $[6, 14, 10]$ & 45 & 1.38 & $[0.34,0.27,0.16,0.11]$ \\
			\hline
		\end{tabular}
	\end{center}
\end{table*}

\begin{figure*}[h]
	\centering
	\begin{subfigure}[t]{0.31\textwidth}
		\centering
		\includegraphics[width=\textwidth]{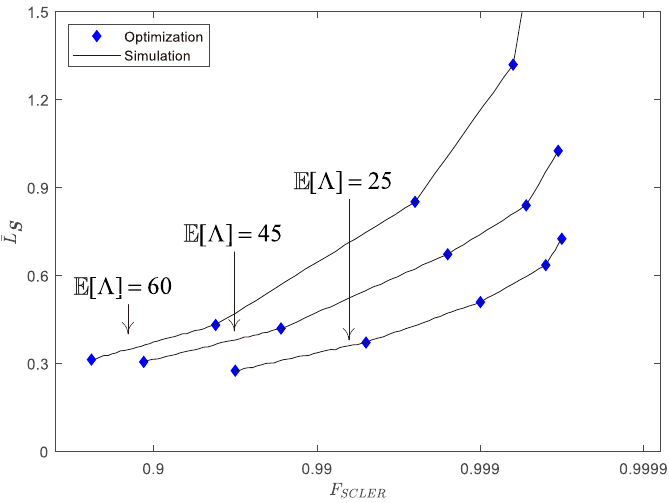}
		\caption{Impact of different ${\mathbb{E}[\Lambda]}$ on the results.}
		\label{result3:sub1}
	\end{subfigure}
	\hfill
	\begin{subfigure}[t]{0.31\textwidth}
		\centering
		\includegraphics[width=\textwidth]{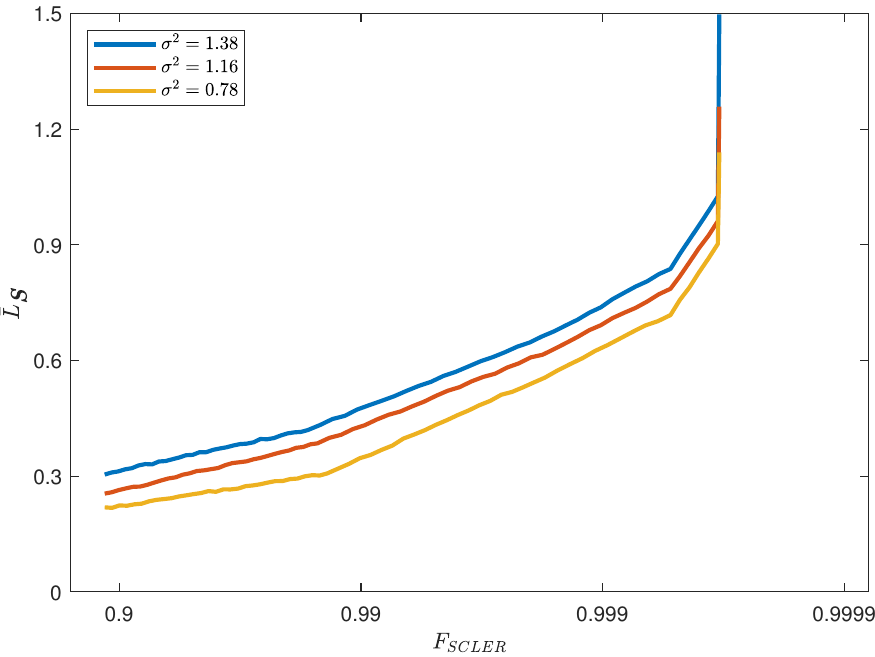}
		\caption{Impact of different $\sigma^2$ on the results.}
		\label{result3:sub2}
	\end{subfigure}
	\hfill
	\begin{subfigure}[t]{0.31\textwidth}
		\centering
		\includegraphics[width=\textwidth]{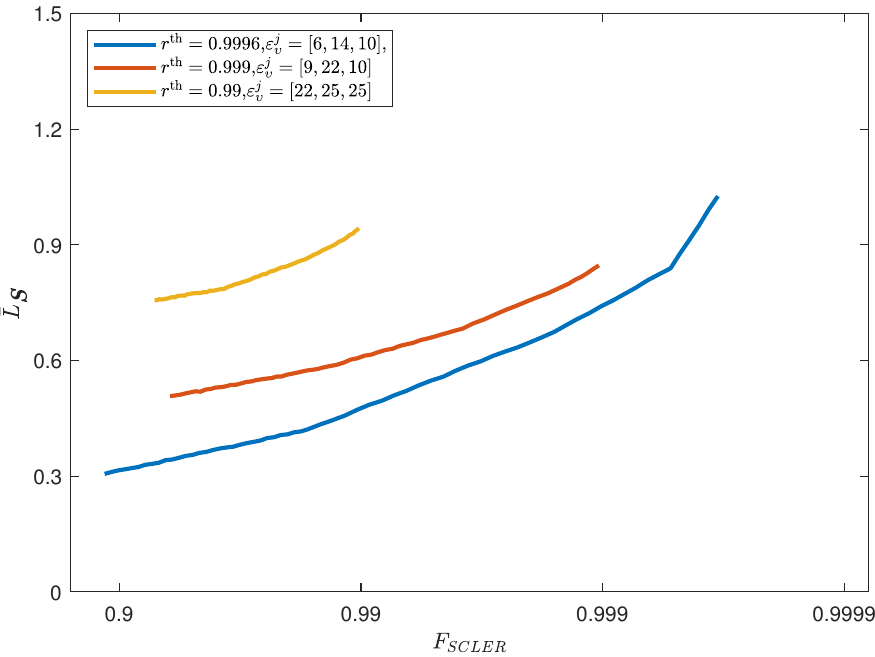}
		\caption{Impact of different $\varepsilon _\upsilon ^j(\%)$ on the results.}
		\label{result3:sub3}
	\end{subfigure}
	\caption{Optimal Delay-Reliability Tradeoff Curves for Various Scenarios.}
	\label{result3}
\end{figure*}

As shown in \ref{result2:sub2}, we have illustrated $F_{SCLER}$ and presented the changes in reliability values under different typical strategies based on Eq.\ref{latency-reliability-security-function}. 
The orange and red lines show $F_{SCLER}$ values that increase with delay, 
$F_{SCLER}$ represents the probability of the system successfully decoding ${\gamma _\upsilon ^j}$ in the legal range, when the allocation weight exceeds the decoding threshold ${\gamma _d}$, the method can be decoded successfully; When ${\gamma _\upsilon ^j}=[0.2\;0.2\;0.9]$, where ${\gamma _3}$ exceeds the safety threshold ${\gamma_s^j}=0.8$, and $F_{SCLER}$ is always zero. We noticed that although NR has the advantage of $t_p=6.25$, however the first non-zero point of the $F_{SCLER}$-curve is still around 10. This is because, under the limit of the safety threshold ${\gamma _s}$, the decoding cannot be completed based on the bearer $Pc$ of single path. Determined by the characteristics of fountain codes, DIEN will be unreliable if and only when all three channels are error-coded, so ${P_e} = 1 - \prod\limits_{j = 1}^{{N_L}} {(1 - P_e^j)} $, the extreme value of $F_{SCLER}$ is 0.9996. The strategy $\emph{S}$ takes effect based on $F_{SCLER}$ \ref{delayaimMCmatrix:1B}, which is one of the foundations of this paper.

%legend('${\gamma _\upsilon ^j}=[0.2\;0.2\;0.9]$', '${\gamma _\upsilon ^j}=[0.6\;0.6\;0.6]$', '${\gamma _\upsilon ^j}=[0.8\;0.8\;0.8]$', 'Location', 'SouthEast','Interpreter', 'latex');

\subsection{Numerical Results}
The SCLER strategy $\emph{S}$ is based on DIEN transmission, and we have analyzed the data protection capabilities in the face of passive attackers in Fig.\ref{decoded-number-versus}. This section will focus on analyzing the optimal trade-off between delay and reliability under the premise of security considerations.

In this subsection, we analyze the network characteristic values based on the basic performance results. We set the maximum transmission block length $B=100$, considering the encoding characteristics, when  $g[t]<100$, random redundant $Ps$ will be added to ensure that ${\gamma _j}{\gamma _\tau } \ge {\gamma _d}$. We consider B5G DIEN containing three PDU sessions, whose transmission characteristics are shown in Fig.\ref{result2:sub1}. The accumulation and arrival of $Ps$ follow the probability distribution ${ {\lambda _{( \cdot )}}} $, and the maximum arrival rate is $N_{\max \lambda} = 300$. For the convenience of analysis, we assume that the minimum granularity of a single timeslot data arrival is 25, namely $ { {{\Lambda _N}} } = { 0,25,50,75,100} $. We sort the PDU sessions as 1-3 according to NR, Wifi, and LTE bearers. $Pc$ erasure at the PDU session layer is considered a separate unreliability other than $Pe$, $\varepsilon [n]$ does not vary with time under the consider of there is no retransmission, and this assumption is reasonable in mobile networks. The number of time slots for each simulation run is ${N_T}=10^5$. Theoretical results are plotted with diamond-shaped dots, while simulation results are marked with lines. The basic simulation parameter settings can be found in Table.\ref{tab2}.

The Fig.\ref{result3:sub1} shows the numerical results of the delay-reliability trade-off. It illustrates that for fixed deletion probabilities of $\varepsilon_\upsilon ^j= [6, 14, 10]$, $F_{SCLER}$ can achieve the minimum average waiting delay while ensuring reliability for average arrival rates ${\mathbb{E}[\Lambda]}$ of 25, 45, and 60. The simulation results match well with the theoretical analysis shown by the diamond-shaped dots, which shows that the derived optimization problem of the threshold-based strategy accurately solves the Eq.\ref{delayaimMC}. The delay-reliability trade-off curve is segmented, which is consistent with the characteristics of the strategy. As the reliability $F{SCLER}$ approaches its extreme value, the delay increases rapidly, and the queue state becomes more unstable. Under the same reliability constraints, as ${\mathbb{E}[\Lambda] }$ increases, the corresponding average queue delay ${\bar {L}_\textbf{\emph{S}}}$ also increases due to the higher throughput of DIEN at this time. In summary, SCLER can achieve a latency-reliability trade-off, and the threshold-based strategy is effective.

In Fig.\ref{result3:sub2}, we use the same setting as in Fig.\ref{result3:sub1} to analyze the delay-reliability trade-off when $\mathbb{E}[\Lambda] = 45$ and $\sigma^2$ =[0.78, 1.16, 1.38]. We observe that the arrival process $\boldsymbol{\lambda}$ with a larger second-order moment $\sigma^2$ has a higher corresponding average queue delay $\bar{L}\textbf{\emph{S}}$, which is longer. As the reliability $F$ approaches $1-P_e$, more delay overhead needs to be added due to the long-tailed distribution of the reliability function. This causes $F_{SCLER}$ to slowly approach the extreme value. In conclusion, the variance has an effect on the reliability-delay trade-off, which helps us to further analyze the threshold-based strategy.

Fig.\ref{result3:sub3} shows the analysis of the delay-reliability trade-off under different deletion probabilities for $\mathbb{E}[\Lambda] = 45$ and three channel transmission state as shown in Table.\ref{tab2}. With the deletion probability increases, a lower code rate is required to achieve the same reliability value. In this case, $\textbf{\emph{S}}$ tends to generate a smaller $g$ to improve the coding redundancy, which increases the queue delay in the DIEN.

\begin{figure}[h]
	\centering
	\begin{subfigure}[t]{0.47\columnwidth}
		\centering
		\includegraphics[width=\textwidth]{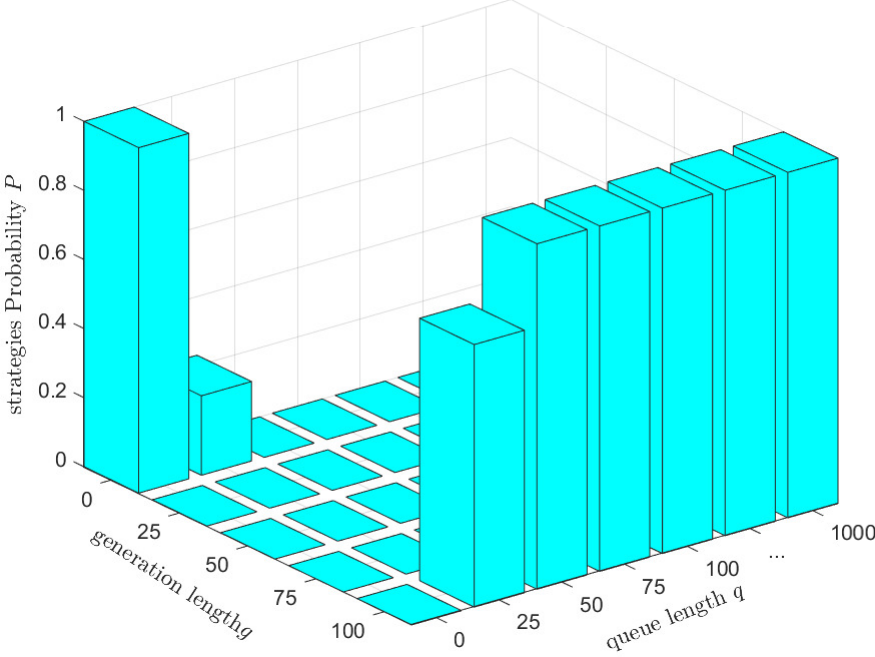}
		\caption{$r^{{\rm{th}}}$= 0.9996}
		\label{result4:sub1}
	\end{subfigure}
	\hfill
	\begin{subfigure}[t]{0.47\columnwidth}
		\centering
		\includegraphics[width=\textwidth]{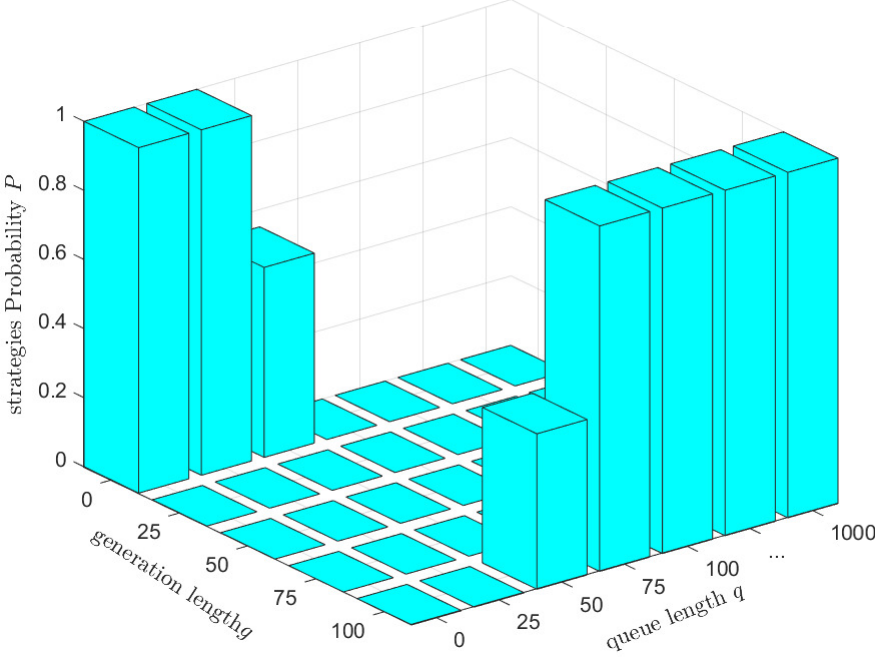}
		\caption{$r^{{\rm{th}}}$= 0.999}
		\label{result4:sub2}
	\end{subfigure}
	\caption{Threshold-based strategy $\textbf{\emph{S}}$, ${\mathbb{E}[\Lambda] }$=45.}
	\label{result4}
\end{figure}

Fig.\ref{result4} shows the specific strategy probability value of the threshold policy. We analyze the change of $\textbf{\emph{S}}$ under different reliability requirements $r^{{\rm{th}}}$ for ${\mathbb{E}[\Lambda] }$=45. The results in Fig.\ref{result3:sub1} have confirmed the effectiveness of the threshold strategy in solving derivative problems. In Fig.\ref{result4:sub1}, the threshold point is at $q_[t]=25$, and the probability of transmission is 0.75, while in Fig.\ref{result4:sub2}, the threshold point is at $q_[t]=50$, and the probability of transmission is 0.47. Considering both reliability and delay, SCLER adjust the threshold policy point to achieve optimal trade-off optimization.

%%%%%%%%%%%%%%%%%%%%%%%%%%%%%%%%%%%%%%%%%%

\section{Conclusion}
This paper investigates how to provide end-to-end low-latency, high-reliability transmission methods between terminal devices and edge nodes in Beyond 5G  edge networks, while considering potential eavesdroppers. We propose a secure and reliable Raptor encoding-based transmission scheme that mitigates the impact of asymmetric delay and bandwidth differences in multi-path transmission, thus enhancing the interaction between paths. We develop a sender-side queue length-aware variable block-length encoding method and implement multi-path split transmission, taking into account a data security threshold that reduces data leakage risks on a larger attack surface in CMT. By analyzing the optimal trade-off decision problem for delay and reliability, we address the constraint optimization problem using a threshold-based strategy. Performance results demonstrate that our approach is effective in achieving the optimization of both delay and reliability, while ensuring data security.

%%%%%%%%%%%%%%%%%%%%%%%%%%%%%%%%%%%%%%%%%%
%% Optional
\begin{appendices}
	\section{Multi-Access PDU session\\Establishment Procedure}
	The current existing mobile network does not yet support the establishment of physically isolated multi-PDU sessions. Therefore, we propose the procedure of establishing diversity interface PDU session suitable for SCLER
	\begin{enumerate}
		\item{In SCLER, both UE and DN can initiate a PDU session establishment request. The corresponding message is PDU Session Establishment Request, which needs to carry the requested DNN (Data Network Name), PDU session type, etc.}
		\item{AMF will send the message to SMF in SBI(service-based interface) format and will report the access type currently supported by UE to the core network.}
		\item{SMF will perform the authentication and authorization of the PDU session, and register on the Unified Data Management (UDM). These contents are not shown in Fig. \ref{SCLER_fig3}. SMF reads the service data of the user subscribed by the user  in the SCLER and invokes the session management policy rule corresponding to the user, which is used for the PDU session QoS control, At this time, SMF has selected the DN and the Anchor UPF.}
		\item{SMF sends PDR to UPF, (PDR is a kind of cell, used to carry policy control information, including rule name, forwarding action, service processing action, etc.), UPF will be assigned with RAN and N3IWF addresses, namely TEID, for RAN and N3IWF uplink data transmission, corresponding to UPF-1 and UPF-2 respectively,}
		\item{SMF requires AMF to relay the SMF's NAS messages ($i.e.$ the session establishment accept the message and the UPF's core network tunnel information for uplink from the RAN) to the UE.}
		\item{AMF initiates a radio resource request to the access point, including 3GPP access and non-3GPP access, and requires relaying NAS messages}
		\item{The NG-RAN and Wifi set up the radio bearer and RAN tunnel after receiving the request, and sending the NAS message to the UE}
		\item{NG-RAN and Wifi use session information (including session id and RAN tunnel) to provide response to AMF}
		\item{AMF relays session information to SMF}
		\item{SMF updates the UPF after receiving the reply and establishes the downlink tunnel from the UPF to the base station}
		\item{Finally, the tunnel from the access point to UPF-A is generated. The UE and UPF-A split and aggregate the data, and the encoding and decoding are also completed in the UE and UPF-A.}
	\end{enumerate}
\end{appendices}

\bibliography{reference}

\begin{thebibliography}{10}
\providecommand{\url}[1]{#1}
\csname url@samestyle\endcsname
\providecommand{\newblock}{\relax}
\providecommand{\bibinfo}[2]{#2}
\providecommand{\BIBentrySTDinterwordspacing}{\spaceskip=0pt\relax}
\providecommand{\BIBentryALTinterwordstretchfactor}{4}
\providecommand{\BIBentryALTinterwordspacing}{\spaceskip=\fontdimen2\font plus
\BIBentryALTinterwordstretchfactor\fontdimen3\font minus
  \fontdimen4\font\relax}
\providecommand{\BIBforeignlanguage}[2]{{%
\expandafter\ifx\csname l@#1\endcsname\relax
\typeout{** WARNING: IEEEtran.bst: No hyphenation pattern has been}%
\typeout{** loaded for the language `#1'. Using the pattern for}%
\typeout{** the default language instead.}%
\else
\language=\csname l@#1\endcsname
\fi
#2}}
\providecommand{\BIBdecl}{\relax}
\BIBdecl

\bibitem{3gpp.21.917}
\BIBentryALTinterwordspacing
3GPP, ``{Summary of Rel-17 Work Items},'' Technical report (TR) 21.917, 01
  2023, version 17.0.1. [Online]. Available:
  \url{https://www.3gpp.org/ftp/Specs/archive/21_series/21.917/}
\BIBentrySTDinterwordspacing

\bibitem{javed2022future}
A.~R. Javed, F.~Shahzad, S.~ur~Rehman, Y.~B. Zikria, I.~Razzak, Z.~Jalil, and
  G.~Xu, ``Future smart cities requirements, emerging technologies,
  applications, challenges, and future aspects,'' \emph{Cities}, vol. 129, p.
  103794, 2022.

\bibitem{navarro2020survey}
J.~Navarro-Ortiz, P.~Romero-Diaz, S.~Sendra, P.~Ameigeiras, J.~J. Ramos-Munoz,
  and J.~M. Lopez-Soler, ``A survey on 5g usage scenarios and traffic models,''
  \emph{IEEE Communications Surveys \& Tutorials}, vol.~22, no.~2, pp.
  905--929, 2020.

\bibitem{li20185g}
Z.~Li, M.~A. Uusitalo, H.~Shariatmadari, and B.~Singh, ``5g urllc: Design
  challenges and system concepts,'' in \emph{2018 15th international symposium
  on wireless communication systems (ISWCS)}.\hskip 1em plus 0.5em minus
  0.4em\relax IEEE, 2018, pp. 1--6.

\bibitem{adhikari20226g}
M.~Adhikari and A.~Hazra, ``6g-enabled ultra-reliable low-latency communication
  in edge networks,'' \emph{IEEE Communications Standards Magazine}, vol.~6,
  no.~1, pp. 67--74, 2022.

\bibitem{zhang2022mobile}
Y.~Zhang and Y.~Zhang, ``Mobile edge computing for beyond 5g/6g,'' \emph{Mobile
  Edge Computing}, pp. 37--45, 2022.

\bibitem{hassan2019edge}
N.~Hassan, K.-L.~A. Yau, and C.~Wu, ``Edge computing in 5g: A review,''
  \emph{IEEE Access}, vol.~7, pp. 127\,276--127\,289, 2019.

\bibitem{hui2021secure}
Y.~Hui, N.~Cheng, Z.~Su, Y.~Huang, P.~Zhao, T.~H. Luan, and C.~Li, ``Secure and
  personalized edge computing services in 6g heterogeneous vehicular
  networks,'' \emph{IEEE Internet of Things Journal}, vol.~9, no.~8, pp.
  5920--5931, 2021.

\bibitem{tedeschi2019edge}
P.~Tedeschi and S.~Sciancalepore, ``Edge and fog computing in critical
  infrastructures: Analysis, security threats, and research challenges,'' in
  \emph{2019 IEEE European Symposium on Security and Privacy Workshops
  (EuroS\&PW)}.\hskip 1em plus 0.5em minus 0.4em\relax IEEE, 2019, pp. 1--10.

\bibitem{yoshizawa2019overview}
T.~Yoshizawa, S.~B.~M. Baskaran, and A.~Kunz, ``Overview of 5g urllc system and
  security aspects in 3gpp,'' in \emph{2019 IEEE Conference on Standards for
  Communications and Networking (CSCN)}.\hskip 1em plus 0.5em minus 0.4em\relax
  IEEE, 2019, pp. 1--5.

\bibitem{chen2019physical}
R.~Chen, C.~Li, S.~Yan, R.~Malaney, and J.~Yuan, ``Physical layer security for
  ultra-reliable and low-latency communications,'' \emph{IEEE Wireless
  Communications}, vol.~26, no.~5, pp. 6--11, 2019.

\bibitem{hamamreh2017ofdm}
J.~M. Hamamreh, E.~Basar, and H.~Arslan, ``Ofdm-subcarrier index selection for
  enhancing security and reliability of 5g urllc services,'' \emph{IEEE
  Access}, vol.~5, pp. 25\,863--25\,875, 2017.

\bibitem{lien20175g}
S.-Y. Lien, S.-L. Shieh, Y.~Huang, B.~Su, Y.-L. Hsu, and H.-Y. Wei, ``5g new
  radio: Waveform, frame structure, multiple access, and initial access,''
  \emph{IEEE communications magazine}, vol.~55, no.~6, pp. 64--71, 2017.

\bibitem{shrivastava20225g}
V.~K. Shrivastava, S.~Baek, and Y.~Baek, ``5g evolution for multicast and
  broadcast services in 3gpp release 17,'' \emph{IEEE Communications Standards
  Magazine}, vol.~6, no.~3, pp. 70--76, 2022.

\bibitem{xie2022optimizing}
Y.~Xie and P.~Ren, ``Optimizing training and transmission overheads for secure
  urllc against randomly distributed eavesdroppers,'' \emph{IEEE Transactions
  on Vehicular Technology}, vol.~71, no.~11, pp. 11\,921--11\,935, 2022.

\bibitem{xu2021quantum}
D.~Xu and P.~Ren, ``Quantum learning based nonrandom superimposed coding for
  secure wireless access in 5g urllc,'' \emph{IEEE Transactions on Information
  Forensics and Security}, vol.~16, pp. 2429--2444, 2021.

\bibitem{farhat2021secure}
J.~Farhat, G.~Brante, R.~D. Souza, and J.~P. Vilela, ``On the secure spectral
  efficiency of urllc with randomly located colluding eavesdroppers,''
  \emph{IEEE Internet of Things Journal}, vol.~8, no.~19, pp. 14\,672--14\,682,
  2021.

\bibitem{3gpp.23.501}
\BIBentryALTinterwordspacing
3GPP, ``{System architecture for the 5G System (5GS)},'' Technical
  Specification (TS) 23.501, 09 2022, version 17.6.0. [Online]. Available:
  \url{https://www.3gpp.org/ftp/Specs/archive/23_series/23.501/}
\BIBentrySTDinterwordspacing

\bibitem{chen2018ultra}
H.~Chen, R.~Abbas, P.~Cheng, M.~Shirvanimoghaddam, W.~Hardjawana, W.~Bao,
  Y.~Li, and B.~Vucetic, ``Ultra-reliable low latency cellular networks: Use
  cases, challenges and approaches,'' \emph{IEEE Communications Magazine},
  vol.~56, no.~12, pp. 119--125, 2018.

\bibitem{suer2019multi}
M.-T. Suer, C.~Thein, H.~Tchouankem, and L.~Wolf, ``Multi-connectivity as an
  enabler for reliable low latency communications—an overview,'' \emph{IEEE
  Communications Surveys \& Tutorials}, vol.~22, no.~1, pp. 156--169, 2019.

\bibitem{li2016multipath}
M.~Li, A.~Lukyanenko, Z.~Ou, A.~Yl{\"a}-J{\"a}{\"a}ski, S.~Tarkoma, M.~Coudron,
  and S.~Secci, ``Multipath transmission for the internet: A survey,''
  \emph{IEEE Communications Surveys \& Tutorials}, vol.~18, no.~4, pp.
  2887--2925, 2016.

\bibitem{ha2019support}
J.~Ha and Y.-I. Choi, ``Support of a multi-access session in 5g mobile
  network,'' in \emph{2019 25th Asia-Pacific Conference on Communications
  (APCC)}.\hskip 1em plus 0.5em minus 0.4em\relax IEEE, 2019, pp. 378--383.

\bibitem{wang2022secure}
J.~Wang and J.~Liu, ``Secure and reliable slicing in 5g and beyond vehicular
  networks,'' \emph{IEEE Wireless Communications}, vol.~29, no.~1, pp.
  126--133, 2022.

\bibitem{cohen2021network}
A.~Cohen, R.~G. D’Oliveira, S.~Salamatian, and M.~M{\'e}dard, ``Network
  coding-based post-quantum cryptography,'' \emph{IEEE Journal on Selected
  Areas in Information Theory}, vol.~2, no.~1, pp. 49--64, 2021.

\bibitem{huang2022cross}
L.~Huang, P.~Ren, and D.~Xu, ``Cross-locking enabled multi-route fountain
  coding for secure transmission,'' in \emph{2022 IEEE 95th Vehicular
  Technology Conference:(VTC2022-Spring)}.\hskip 1em plus 0.5em minus
  0.4em\relax IEEE, 2022, pp. 1--5.

\bibitem{luby2002lt}
M.~Luby, ``Lt codes,'' in \emph{The 43rd Annual IEEE Symposium on Foundations
  of Computer Science, 2002. Proceedings.}\hskip 1em plus 0.5em minus
  0.4em\relax IEEE Computer Society, 2002, pp. 271--271.

\bibitem{he2021delay}
M.~He, C.~Hua, W.~Xu, P.~Gu, and X.~S. Shen, ``Delay optimal concurrent
  transmissions with raptor codes in dual connectivity networks,'' \emph{IEEE
  Transactions on Network Science and Engineering}, vol.~8, no.~2, pp.
  1478--1491, 2021.

\bibitem{cui2014fmtcp}
Y.~Cui, L.~Wang, X.~Wang, H.~Wang, and Y.~Wang, ``Fmtcp: A fountain code-based
  multipath transmission control protocol,'' \emph{IEEE/ACM Transactions on
  Networking}, vol.~23, no.~2, pp. 465--478, 2014.

\bibitem{luby2007raptor}
M.~Luby, A.~Shokrollahi, M.~Watson, and T.~Stockhammer, ``Raptor forward error
  correction scheme for object delivery,'' Tech. Rep., 2007.

\bibitem{abbas2019performance}
R.~Abbas, M.~Shirvanimoghaddam, T.~Huang, Y.~Li, and B.~Vucetic, ``Performance
  analysis of short analog fountain codes,'' in \emph{2019 IEEE Globecom
  Workshops (GC Wkshps)}.\hskip 1em plus 0.5em minus 0.4em\relax IEEE, 2019,
  pp. 1--6.

\bibitem{jain2020rateless}
S.~Jain and R.~Bose, ``Rateless-code-based secure cooperative transmission
  scheme for industrial iot,'' \emph{IEEE Internet of Things Journal}, vol.~7,
  no.~7, pp. 6550--6565, 2020.

\bibitem{yi2014achieving}
M.~Yi, X.~Ji, K.~Huang, H.~Wen, and B.~Wu, ``Achieving strong security based on
  fountain code with coset pre-coding,'' \emph{IET Communications}, vol.~8,
  no.~14, pp. 2476--2483, 2014.

\bibitem{GUO2023109716}
\BIBentryALTinterwordspacing
Z.~Guo, X.~Ji, W.~You, M.~Xu, Y.~Zhao, Z.~Cheng, and D.~Zhou, ``Delay optimal
  for reliability-guaranteed concurrent transmissions with raptor code in
  multi-access 6g edge network,'' \emph{Computer Networks}, p. 109716, 2023.
  [Online]. Available:
  \url{https://www.sciencedirect.com/science/article/pii/S1389128623001615}
\BIBentrySTDinterwordspacing

\bibitem{kwon2014mpmtp}
O.~C. Kwon, Y.~Go, Y.~Park, and H.~Song, ``Mpmtp: Multipath multimedia
  transport protocol using systematic raptor codes over wireless networks,''
  \emph{IEEE Transactions on Mobile Computing}, vol.~14, no.~9, pp. 1903--1916,
  2014.

\bibitem{nielsen2017ultra}
J.~J. Nielsen, R.~Liu, and P.~Popovski, ``Ultra-reliable low latency
  communication using interface diversity,'' \emph{IEEE Transactions on
  Communications}, vol.~66, no.~3, pp. 1322--1334, 2017.

\bibitem{3gpp.29.244}
\BIBentryALTinterwordspacing
3GPP, ``{Interface between the Control Plane and the User Plane Nodes},''
  Technical Specification (TS) 29.244, 12 2022, version 18.0.1. [Online].
  Available: \url{https://www.3gpp.org/ftp/Specs/archive/29_series/29.244/}
\BIBentrySTDinterwordspacing

\bibitem{jadin2017securing}
M.~Jadin, G.~Tihon, O.~Pereira, and O.~Bonaventure, ``Securing multipath tcp:
  Design \& implementation,'' in \emph{IEEE INFOCOM 2017-IEEE Conference on
  Computer Communications}.\hskip 1em plus 0.5em minus 0.4em\relax IEEE, 2017,
  pp. 1--9.

\bibitem{apiecionek2015multi}
{\L}.~Apiecionek, M.~Sobczak, W.~Makowski, and T.~Vince, ``Multi path
  transmission control protocols as a security solution,'' in \emph{2015 IEEE
  13th International Scientific Conference on Informatics}.\hskip 1em plus
  0.5em minus 0.4em\relax IEEE, 2015, pp. 27--31.

\bibitem{liu2022secure}
J.~Liu, J.~Xu, S.~Li, X.~Cui, and Y.~Zhang, ``A secure multi-path transmission
  algorithm based on fountain codes,'' \emph{Transactions on Emerging
  Telecommunications Technologies}, vol.~33, no.~5, p. e4450, 2022.

\bibitem{noura2022network}
H.~N. Noura, R.~Melki, and A.~Chehab, ``Network coding and mptcp: Enhancing
  security and performance in an sdn environment,'' \emph{Journal of
  Information Security and Applications}, vol.~66, p. 103165, 2022.

\bibitem{yang2001improving}
J.~Yang and S.~Papavassiliou, ``Improving network security by multipath traffic
  dispersion,'' in \emph{2001 MILCOM Proceedings Communications for
  Network-Centric Operations: Creating the Information Force (Cat. No.
  01CH37277)}, vol.~1.\hskip 1em plus 0.5em minus 0.4em\relax IEEE, 2001, pp.
  34--38.

\bibitem{singh2015survey}
S.~K. Singh, T.~Das, and A.~Jukan, ``A survey on internet multipath routing and
  provisioning,'' \emph{IEEE Communications Surveys \& Tutorials}, vol.~17,
  no.~4, pp. 2157--2175, 2015.

\bibitem{kuo2008dynamic}
C.-F. Kuo, A.-C. Pang, and S.-K. Chan, ``Dynamic routing with security
  considerations,'' \emph{IEEE transactions on parallel and distributed
  systems}, vol.~20, no.~1, pp. 48--58, 2008.

\bibitem{3gpp.23.502}
\BIBentryALTinterwordspacing
3GPP, ``{Procedures for the 5G System (5GS)},'' Technical Specification (TS)
  23.502, 09 2022, version 17.6.0. [Online]. Available:
  \url{https://www.3gpp.org/ftp/Specs/archive/23_series/23.502/}
\BIBentrySTDinterwordspacing

\bibitem{julio2020r}
Y.~R. Julio, I.~G. Garcia, and J.~Marquez, ``R-iot: An architecture based on
  recoding rlnc for iot wireless network with erase channel,'' in
  \emph{Information Technology and Systems: Proceedings of ICITS 2020}.\hskip
  1em plus 0.5em minus 0.4em\relax Springer, 2020, pp. 579--588.

\bibitem{yang2022network}
S.~Yang and R.~W. Yeung, ``Network communication protocol design from the
  perspective of batched network coding,'' \emph{IEEE Communications Magazine},
  vol.~60, no.~1, pp. 89--93, 2022.

\bibitem{boualouache2022federated}
A.~Boualouache and T.~Engel, ``Federated learning-based scheme for detecting
  passive mobile attackers in 5g vehicular edge computing,'' \emph{Annals of
  Telecommunications}, pp. 1--20, 2022.

\bibitem{huang2017coding}
W.~Huang, ``Coding for security and reliability in distributed systems,'' Ph.D.
  dissertation, California Institute of Technology, 2017.

\bibitem{shokrollahi2006raptor}
A.~Shokrollahi, ``Raptor codes,'' \emph{IEEE transactions on information
  theory}, vol.~52, no.~6, pp. 2551--2567, 2006.

\bibitem{karp2004finite}
R.~Karp, M.~Luby, and A.~Shokrollahi, ``Finite length analysis of lt codes,''
  in \emph{International Symposium onInformation Theory, 2004. ISIT 2004.
  Proceedings.}\hskip 1em plus 0.5em minus 0.4em\relax IEEE, 2004, p.~39.

\bibitem{shokrollahi2009theory}
A.~Shokrollahi, ``Theory and applications of raptor codes,'' \emph{Proceedings
  of MathKnow}, pp. 59--89, 2009.

\bibitem{wang2016performance}
P.~Wang, G.~Mao, Z.~Lin, M.~Ding, W.~Liang, X.~Ge, and Z.~Lin, ``Performance
  analysis of raptor codes under maximum likelihood decoding,'' \emph{IEEE
  Transactions on Communications}, vol.~64, no.~3, pp. 906--917, 2016.

\bibitem{strom20155g}
E.~G. Str{\"o}m, P.~Popovski, and J.~Sachs, ``5g ultra-reliable vehicular
  communication,'' \emph{arXiv preprint arXiv:1510.01288}, 2015.

\bibitem{wang2019event}
F.~Wang, G.~Wen, Z.~Peng, T.~Huang, and Y.~Yu, ``Event-triggered consensus of
  general linear multiagent systems with data sampling and random packet
  losses,'' \emph{IEEE Transactions on Systems, Man, and Cybernetics: Systems},
  vol.~51, no.~2, pp. 1313--1321, 2019.

\bibitem{khatua2020application}
P.~K. Khatua, V.~K. Ramachandaramurthy, P.~Kasinathan, J.~Y. Yong,
  J.~Pasupuleti, and A.~Rajagopalan, ``Application and assessment of internet
  of things toward the sustainability of energy systems: Challenges and
  issues,'' \emph{Sustainable Cities and Society}, vol.~53, p. 101957, 2020.

\end{thebibliography}
\bibliographystyle{IEEEtran}
\end{document}